%% file: main.tex
\RequirePackage{fix-cm}
\documentclass[twocolumn,epjc3,twoside]{svjour3}

\makeatletter
\let\cl@chapter\undefined
\makeatletter

\smartqed  %
\RequirePackage{graphicx}
\usepackage[aboveskip=1pt,belowskip=0pt,justification=centering,hypcap=true]{subcaption}
\usepackage{tocvsec2}
\usepackage{pdflscape}
\usepackage{amsfonts,amssymb}
\usepackage{amsmath,bbm}
\usepackage{graphicx}
\usepackage{colonequals}
\usepackage[unskipbreak]{cuted}
\usepackage[shortlabels]{enumitem}
\usepackage{makecell}
\usepackage{multirow}
\usepackage{xcolor}
\usepackage{dpfloat}
\usepackage{xspace}
\usepackage{textcomp}
\usepackage{hyperref}
\usepackage{rotating}
\usepackage{cite}
\usepackage{etoolbox}
\usepackage{bm}
\usepackage{upgreek}
\usepackage[normalem]{ulem}
\usepackage{adjustbox} %
\hypersetup{
    colorlinks,
    linkcolor={red!50!black},
    citecolor={blue!50!black},
    urlcolor={blue!80!black}
}
\usepackage[T1]{fontenc}
\usepackage[capitalise]{cleveref}
\usepackage{adjustbox}

\usepackage{footmisc}

\crefname{section}{Sec.}{Secs.}
\Crefname{section}{Section}{Sections}

\crefname{subsection}{Sec.}{Secs.}
\Crefname{subsection}{Section}{Sections}

\definecolor{decompred}{rgb}{1, 0, 0}
\definecolor{decompgreen}{rgb}{0.0, 0.5, 0.0}
\definecolor{decompyellow}{rgb}{0.98, 0.63, 0.0}
\definecolor{decompblue}{rgb}{0, 0, 1}
\definecolor{decompgrey}{rgb}{0.5, 0.5, 0.5}

\crefformat{appendix}{#2#1#3}
\makeatletter
\def\@mkboth#1#2{}
\newlength\appendixwidth
\preto\appendix{\addtocontents{toc}{\protect\patchl@section}}
\newcommand{\patchl@section}{%
  \settowidth{\appendixwidth}{\textbf{Appendix }}%
  \addtolength{\appendixwidth}{1.5em}%
  \patchcmd{\l@section}{1.5em}{\appendixwidth}{}{}%
}
\makeatother

\DeclareRobustCommand{\mposd}{\ensuremath{\textsc{Mpos}{\delta}}\xspace}
\DeclareRobustCommand{\mpos}{\ensuremath{\textsc{Mpos}}\xspace}
\DeclareRobustCommand{\pos}{\ensuremath{\textsc{Pos}}\xspace}

\DeclareRobustCommand{\krk}{\ensuremath{\textsc{Krk}}\xspace}

\DeclareRobustCommand{\dis}{\ensuremath{\textsc{Dis}}\xspace}
\DeclareRobustCommand{\phys}{\ensuremath{\textsc{Phys}}\xspace}
\DeclareRobustCommand{\aversa}{\ensuremath{\textsc{Aversa}}\xspace}
\DeclareRobustCommand{\gen}{\ensuremath{\mathrm{GEN}}\xspace}

\DeclareRobustCommand{\rFS}{\ensuremath{\mathrm{FS}}\xspace}

\DeclareRobustCommand{\alphas}{\alpha_{\mathrm{s}}}

\DeclareRobustCommand{\msbar}{\ensuremath{{\overline{\mathrm{MS}}}}\xspace}
\DeclareRobustCommand{\order}[1]{\mathcal{O}\!\left(#1\right)}
\DeclareRobustCommand{\muf}{\mu_{\rF}}

\DeclareMathOperator{\Tr}{Tr}

\DeclareRobustCommand{\dd}{\ensuremath{\mathrm{d}}}
\DeclareRobustCommand{\rF}{\ensuremath{\mathrm{F}}}
\DeclareRobustCommand{\rK}{\ensuremath{\mathrm{K}}}
\DeclareRobustCommand{\rP}{\ensuremath{\mathrm{P}}}

\DeclareRobustCommand{\pqq}{\ensuremath{p_{qq}(z)}}
\DeclareRobustCommand{\pqg}{\ensuremath{p_{qg}(z)}}
\DeclareRobustCommand{\pgq}{\ensuremath{p_{gq}(z)}}
\DeclareRobustCommand{\pgg}{\ensuremath{p_{gg}(z)}}

\DeclareRobustCommand{\ensuremathrm}[1]{\ensuremath{\mathrm{#1}}\xspace}

\DeclareRobustCommand{\rF}{\ensuremathrm{F}}

\DeclareRobustCommand{\rK}{\ensuremathrm{K}}

\DeclareRobustCommand{\rT}{\ensuremathrm{T}}

\DeclareRobustCommand{\rFS}{\mathrm{FS}}

\DeclareRobustCommand{\alphas}{\alpha_{\mathrm{s}}}

\DeclareRobustCommand{\nf}{\ensuremath{n_{f}}}
\DeclareRobustCommand{\cf}{\ensuremath{C_{F}}}
\DeclareRobustCommand{\ca}{\ensuremath{C_{A}}}

\DeclareRobustCommand{\tr}{\ensuremath{T_{R}}}
\DeclareRobustCommand{\tR}{\ensuremath{T_{R}}}

\DeclareRobustCommand{\egamma}{\ensuremathrm{\gamma_{\mathrm{E}}}}

\DeclareRobustCommand{\dd}{\ensuremath{\mathrm{d}}}

\DeclareRobustCommand{\pqq}{p_{qq}(z)}
\DeclareRobustCommand{\pqg}{p_{qg}(z)}
\DeclareRobustCommand{\pgq}{p_{gq}(z)}
\DeclareRobustCommand{\pgg}{p_{gg}(z)}

\DeclareMathOperator{\li2}{Li_2}

\newcommand{\MT}{\texttt{MT}\xspace}
\newcommand{\HPL}{\texttt{HPL}\xspace}
\newcommand{\Mathematica}{\textsf{Mathematica}\xspace}

\mathchardef\mhyphen="2D

\DeclareRobustCommand{\nlo}{\ensuremath{\text{NLO}}\xspace}

\newcommand{\krknlo}{{\textsf{KrkNLO}}\xspace}

\journalname{Eur. Phys. J. C}

\begin{document}

\title{PDF evolution in alternative factorisation schemes}
\author{S. Delorme\thanksref{e1,addr1}
        \and
        A. Kusina\thanksref{e2,addr2} 
        \and
        A. Si\'odmok\thanksref{e3,addr3} 
        \and
        J. Whitehead\thanksref{e4,addr3}}
\thankstext{e1}{e-mail: stephane.delorme@us.edu.pl}
\thankstext{e2}{e-mail: aleksander.kusina@ifj.edu.pl}
\thankstext{e3}{e-mail: andrzej.siodmok@uj.edu.pl}
\thankstext{e4}{e-mail: james.whitehead@uj.edu.pl}

\institute{Institute of Physics, University of Silesia, Katowice, Poland \label{addr1}
\and
Institute of Nuclear Physics, Polish Academy of Sciences,
31-342 Kraków, Poland \label{addr2}
\and
Jagiellonian University, ul.\ prof.\ Stanisława Łojasiewicza 11, 30-348 Kraków, Poland \label{addr3}
}

\preprintnumbers{IFJPAN-IV-2026-14/MCNET-26-16}

\date{Submitted: / Revised: }

\onecolumn

{\setlength{\parindent}{0pt}\maketitle}
\setlength{\parindent}{15pt}

\begin{abstract}

Beyond leading order, parton distribution functions (PDFs) require
a choice of factorisation scheme to be defined unambiguously.
Different choices of factorisation scheme lead to PDFs that satisfy
modified DGLAP evolution equations, relative to the conventional \msbar scheme.
In this paper we derive the NLO DGLAP splitting functions for
PDFs in alternative factorisation schemes,
including for a parametrised scheme spanning a subspace of
the general factorisation-scheme space.
We plot their Mellin-space counterparts, the anomalous dimensions,
and study the leading large- and small-$x$ behaviour, relevant to resummation.
We find that the leading behaviour admits a natural interpretation as a
modified effective evolution scale.
This is an essential step towards being able to evolve and
fit PDFs in alternative schemes for use within QCD calculations.


\end{abstract}

\tableofcontents

\vfil

\section{Introduction}
\label{intro}

Beyond leading order, the factorisation of hadronic cross-sections into the convolution of
a perturbative `hard' partonic cross-section with
universal parton distribution functions (PDFs)
necessitates a choice of `factorisation scheme' (FS),
identifying which terms to consider as universal, and which process-specific.
PDFs in different schemes are related to each other by convolution
with a transformation kernel, defined
perturbatively up to the relevant order in $\alphas$.

The evolution of PDFs with the so-called `factorisation scale', $\muf$, is governed by a system of coupled integro-differential equations, the DGLAP equations
\cite{Dokshitzer:1977sg,Gribov:1972ri,Altarelli:1977zs}.
The DGLAP equations are scheme-dependent, and performing
scale-evolution by solving them introduces formally higher-order terms.
As a result, a purely perturbative transformation of
PDFs evolved in one scheme to another does not capture
the higher-order differences (all-order resummation effects) which are present when evolving PDFs directly in a given scheme.

While early calculations favoured the DIS scheme \cite{Altarelli:1978id},
all modern calculations use the modified-minimal subtraction (\msbar) scheme \cite{Bardeen:1978yd} for its simplicity in fixed-order calculations.
Early global PDF fitting collaborations provided PDFs in both schemes,
but the rapid theoretical and computational progress of the NNLO revolution
has led to all PDFs,
as well as the calculations and numerical tools required to fit PDFs
and calculate predictions for hadronic collisions,
being available only in the \msbar scheme.

As a consequence, the theoretical uncertainty of perturbative QCD predictions
attributable to the choice of the \msbar scheme is unknown.
This uncertainty, factorisation scheme uncertainty, is formally very similar
to factorisation scale uncertainty: both arise from formally higher-order
terms which, in certain kinematic limits, may be large despite their formal suppression.
In particular, it may be expected that the role of logarithmic contributions,
both locally in scale at the level of the hard process,
and cumulatively in the DGLAP evolution of PDFs, could be significant.

This was exploited in early work on threshold resummation \cite{Sterman:1986aj},
using the factorisation scheme freedom to identify the all-orders logarithmic behaviour and render it amenable to resummation.
Subsequently these logarithmic contributions have been studied further, primarily through the lens of the
\msbar scheme,
in both the large-$x$ (`soft gluon') limit
\cite{Catani:1989ne,Catani:1996yz}
and the small-$x$ (`high-energy gluon') limit
\cite{Catani:1994sq,Ball:1995vc,Ball:1999sh,Altarelli:2005ni,Bonvini:2016wki,Bonvini:2017ogt,Bonvini:2018xvt,Bonvini:2026cxp}.
This includes the construction of PDF sets
that include resummation effects
\cite{Corcella:2005us,Bonvini:2015ira,Bonvini:2017ogt,Ball:2017otu,xFitterDevelopersTeam:2018hym}.
A complementary approach attempts to sidestep
the question of unphysical scheme-dependence entirely, using physical distributions in place of PDFs, and 
expressing cross-sections in terms of other observable distributions \cite{Furmanski:1981cw,Catani:1996sc,Moch:2009hr,Hentschinski:2013zaa,Lappi:2023lmi,Lappi:2024dvv}.

Recently, new factorisation schemes have been proposed which could offer concrete calculational
advantages\cite{Jadach:2015mza,Jadach:2016acv,Jadach:2016qti,Candido:2020yat,Candido:2023ujx,Aversa:1988vb,Oliveira:2013aug,Ramalho:1983aq,Sterman:1986aj,Nagy:2016pwq,Nagy:2017ggp,Nagy:2022bph}, such as the \krk scheme~\cite{Jadach:2016qti} required by the \krknlo method~\cite{Jadach:2015mza,Jadach:2016qti,Sarmah:2024hdk,Sarmah:2025vnb} for parton shower matching.

Once a factorisation scheme has been chosen,
there are three inequivalent methods of obtaining a PDF set which is formally in a given factorisation
scheme:%
\footnote{Here we tacitly assume that PDFs are determined through global fits of phenomenological predictions to experimental data, as is generally the case for present-day phenomenological studies. However, in recent years there have been considerable developments in obtaining PDFs directly from lattice QCD~\cite{Lin:2017snn,Cichy:2018mum,Constantinou:2020hdm},
where factorisation-scheme freedom also enters \cite{Martinelli:1994ty,Ma:2014jla,Ma:2017pxb,Stewart:2017tvs}.}
\begin{enumerate}[(i)]
    \item by direct fitting in the scheme;
    \item by transforming a PDF fitted in another scheme at a single scale,
    and performing DGLAP evolution to other scales in the new scheme;
    \item by transforming a PDF, fitted and DGLAP-evolved in another scheme, locally at each scale, as in \cite{Delorme:2025teo}.
\end{enumerate}
Making use of alternative schemes via (i) or (ii)
requires the generalisation beyond the \msbar scheme
of the DGLAP equations and the associated evolution (and for (i), PDF fitting) infrastructure.

In this paper, we begin to address (ii),
by calculating the NLO DGLAP splitting functions for two concrete schemes of interest,
the \krk \cite{Jadach:2015mza,Jadach:2016acv,Jadach:2016qti,Jadach:2011cr,Sarmah:2024hdk,Sarmah:2025vnb}
and \phys \cite{Oliveira:2013aug} schemes,
and for a parametrised family of schemes as introduced in \cite{Delorme:2025teo}
to explore the scheme-variation envelope.
This calculation is readily adaptable to any other scheme of interest.
It is also a prerequisite for (i).

The paper is organised as follows. In \cref{sec:facSchemes} we introduce our notation and conventions,
and recapitulate the transformation relations between different schemes.
In \cref{sec:calcres} we summarise the calculation and present our results, the splitting functions in alternative schemes in momentum space,
and in \cref{sec:anomalousdims} the associated plots of their Mellin-transforms, i.e.\ the corresponding anomalous dimensions.
In \cref{sec:asymptotics} we consider the behaviour of the general
case in the large-$z$ and small-$z$ limits,
and its implications for the threshold and small-$x$ resummation programmes
in a general factorisation scheme. Finally, in \cref{sec:conclusions} we summarise our results.

\section{Factorisation schemes}
\label{sec:facSchemes}

We follow the notation and conventions of \cite{Delorme:2025teo},
which we summarise here and in \cref{sec:app_conventions}.
We define PDFs in different factorisation schemes according to their relationship
with PDFs in the \msbar scheme, via a transformation of the form
\begin{align}
	\label{eq:fstransformations_faFSvec}
	\mathbf{f}^{\rFS}
	&=
	\mathbb{K}^{\msbar \to \rFS}
	\otimes
	\mathbf{f}^{\msbar},
\end{align}
where explicitly
\begin{align}
	\label{eq:fstransformations_faFS}
	f^{\rFS}_a (x, \mu)
	&=
	\sum_b
	\int_x^1
	\frac{\dd z}{z}
	\;
	\mathbb{K}^{\msbar \to \rFS}_{ab} \left(z, \mu \right) \ 
	f_b^{\msbar} \left(\frac{x}{z}, \mu \right),
\end{align}
performed locally for each scale $\mu$.%
\footnote{The PDF here is the leading-twist object
defined within collinear factorisation, so higher-twist, power-suppressed terms are not included on either side of
\cref{eq:fstransformations_faFS}.}

\subsection{DGLAP evolution}
\label{subsec:fsdglap}

DGLAP evolution~\cite{Altarelli:1977zs,Gribov:1972ri,Dokshitzer:1977sg}
holds in other factorisation schemes, with modified splitting functions $\mathbb{P}^{\rFS}$:
\begin{align}
	\label{eq:dglap}
	\mu^2 \frac{\partial}{\partial \mu^2} \mathbf{f}^{\rFS}(\mu)
	  = {} &
    \mathbb{P}^{\rFS} (\mu) 
	\otimes
	\mathbf{f}^{\rFS} (\mu) .
\end{align}

The relationship between the splitting functions in factorisation scheme FS and \msbar{} 
can be obtained upon taking the logarithmic derivative $\partial / \partial (\log \mu^2)$
of \cref{eq:fstransformations_faFSvec}. 
\begin{align}
	\label{eq:dglapkernels}
	\mathbb{P}^{\rFS} (\mu)
	= {} &
    \mu^2 \left( \frac{\partial}{\partial \mu^2} 
    \mathbb{K}^{\msbar \to \rFS} (\mu) \right)
    \otimes
    \mathbb{K}^{\rFS \to \msbar} (\mu)
    + 
    \mathbb{K}^{\msbar \to \rFS} (\mu)
    \otimes
    \mathbb{P}^{\msbar} (\mu) 
    \otimes
    \mathbb{K}^{\rFS \to \msbar} (\mu) ,
\end{align}
or using the perturbative expansion of \cref{eq:fstransformations_pertexp}
and expanding according to the convention of \cref{eq:pertexpconv},
\begin{align}
    \rP^{\rFS(0)}_{ab}
    & {} =
    \rP^{\msbar(0)}_{ab} \equiv 0 ,
    \\
    \rP^{\rFS(1)}_{ab}
    & {} =
    \rP^{\msbar(1)}_{ab}
    + \mu^2 \frac{\partial}{\partial \mu^2} 
    \rK_{ab}^{\msbar \to \rFS} (\mu) ,
    \\ \label{eq:PFS2ab}
    \rP^{\rFS(2)}_{ab}
    & {} =
    \rP^{\msbar(2)}_{ab}
    + \mu^2 \frac{\partial}{\partial \mu^2} 
    \rK_{ab}^{\msbar \to \rFS(2)} (\mu)  
    - \sum_c \left( \mu^2 \frac{\partial}{\partial \mu^2} 
    \rK_{ac}^{\msbar \to \rFS} (\mu)\right) \otimes \rK_{cb}^{\msbar \to \rFS} (\mu)
    + \frac{\beta^{(2)}}{2\pi} \rK_{ab}^{\msbar \to \rFS} (\mu) 
    \\ \notag
    & {}
    + \sum_c \Bigl(
    \rK_{ac}^{\msbar \to \rFS} (\mu) \otimes \rP^{\msbar(1)}_{cb}
    -
    \rP^{\msbar(1)}_{ac} \otimes \rK_{cb}^{\msbar \to \rFS} (\mu)
    \Bigr).
\end{align}
The modifications to the DGLAP kernels therefore arise
in turn from
any explicit scale-dependence of the transformation
kernels $\rK^{\msbar\to\rFS}(z;\mu)$,
the QCD $\beta$-function,
and the \msbar DGLAP kernels of the input partons.

Considering the transformation kernels and splitting functions as matrices in flavour space,
we can identify the difference of convolutions in the NLO transformation as the component-wise entry of a commutator,
with Mellin convolution as the product,
\begin{align}
\label{eq:DGLAPFStrans_comm}
\bigl[\mathbb{K}^{(1)}, \mathbb{P}^{(1)}\bigr]_{ab}  {} =
(\mathbb{K}^{(1)} \otimes \mathbb{P}^{(1)} - \mathbb{P}^{(1)} \otimes \mathbb{K}^{(1)})_{ab}
    = \sum_c \Bigl(
    \mathrm{K}^{(1)}_{ac} \otimes \mathrm{P}^{(1)}_{cb}
    -
    \mathrm{P}^{(1)}_{ac} \otimes \mathrm{K}^{(1)}_{cb}
    \Bigr).
\end{align}

For the NLO factorisation-scheme transformations considered in \cite{Delorme:2025teo},
there is no explicit dependence within $\rK$ on $\mu$, and so
we suppress the possible explicit $\mu$-dependence of
$\rK^{\msbar \to \rFS}_{ab}(z, \mu)$.
In this case PDFs in alternative schemes obey a
DGLAP evolution equation modified only at NLO and higher orders.
Concretely,
for the schemes considered here, the DGLAP kernels
may be calculated as
\begin{align}
    \rP^{\rFS(1)}_{ab}
    & {} =
    \rP^{\msbar(1)}_{ab}
    \\ \label{eq:DGLAPFStrans}
    \rP^{\rFS(2)}_{ab}
    & {} =
    \rP^{\msbar(2)}_{ab}
    + \frac{\beta^{(2)}}{2\pi} \rK_{ab}^{\msbar \to \rFS}
    + \bigl[\mathbb{K}^{\msbar \to \rFS(1)}, \mathbb{P}^{\msbar(1)}\bigr]_{ab}.
\end{align}
For brevity we drop the $\rFS$ label at leading order, since the leading-order splitting functions are scheme-independent.
Note that here, when interpreting the indices,
\begin{align}
    \rK_{qg}^{\msbar \to \rFS} = 2\nf \rK_{q_ig}^{\msbar \to \rFS}.
\end{align}

Expressions for the spacelike \nlo kernels in the \msbar scheme
$\rP^{\msbar(2)}_{ab} (z)$
are available in the literature,
for example in
\cite{Ellis:1996mzs}.
Here we provide
$\rP^{\rFS(2)}_{q_i q_i},$
$\rP^{\rFS(2)}_{q_i q_k},$
$\rP^{\rFS(2)}_{q_i \overline{q}_i},$
$\rP^{\rFS(2)}_{q_i \overline{q}_k},$
$\rP^{\rFS(2)}_{q_ig},$
$\rP^{\rFS(2)}_{gq},$
$\rP^{\rFS(2)}_{gg}$
where $k \neq i$,  for the considered alternative factorisation schemes.
The kernels for the singlet--gluon coupled evolution may be obtained from these as
\begin{align}
    \rP^{\rFS(k)}_{qg} &{} = 2 \nf \rP^{\rFS(k)}_{q_ig} , \\
    \rP^{\rFS(k)}_{qq} &{} = 
    \rP^{\rFS(k)}_{q_iq_i}
    + \rP^{\rFS(k)}_{q_i\overline{q}_i}
    + (\nf-1) \left(\rP^{\rFS(k)}_{q_iq_k} + \rP^{\rFS(k)}_{q_i\overline{q}_k}\right).
\end{align}

Although we focus throughout on initial-state (`spacelike') DGLAP evolution as appropriate for PDFs,
the final-state (`timelike') splitting functions for fragmentation functions are identical at leading-order
(`Gribov--Lipatov reciprocity' \cite{Gribov:1972rt}),
and differ at NLO by a known shift, given in \cite{Furmanski:1980cm}.
As a consequence, the transformation relation \cref{eq:PFS2ab} remains valid for
timelike evolution, and so the spacelike results presented here can be adapted for timelike
evolution by adding the same spacelike-to-timelike transition function used for the \msbar scheme.%
\footnote{However, factorisation schemes defined for hadronic collisions
may not have the same physical interpretation when applied directly to final-state evolution.}

\subsection{Scheme definitions}
\label{subsec:definitions}
We decompose convolution kernels as arise up to NLO in QCD as
\begin{align}
\label{eq:tableDecomp}
	K(z) =  {} & 
	\sum_{k=0}^1 a_k \, \mathcal{D}_k (z) 
	+ b(z) \log (1-z)
	+ c(z) \log z
	+ P(z)
	- \Delta \, \delta(1-z),
\end{align}
where $b(z), c(z), P(z)$ are rational
functions and, concretely, for all the kernels we consider, 
can be expressed as Laurent series with at most a simple pole at $z=0$ and otherwise regular on $(0,1]$.
The distributions $\mathcal{D}_k(z)$
are defined as
\begin{equation}
    \label{eq:Dk_kernels}
	\mathcal{D}_k(z) = \left[ \frac{\log^k (1-z)}{1-z} \right]_+
\end{equation}
where the `plus-distribution' regularisation,
for functions otherwise singular at $z=1$,
is defined by its action upon integration against
a smooth function $f$ as
\begin{align}
    \int_x^1 \dd z \, f(z) \left[ \frac{g(z)}{1-z} \right]_+
    & {} =
    \int_x^1 \dd z \, \frac{f(z) - f(1)}{1 - z} \, g(z)
    - f(1) \int_0^x \dd z \, \frac{g(z)}{1-z}.
\end{align}
The plus distribution $\mathcal{D}_1$ is the most singular that arises in 
coefficient functions at NLO in QCD.

\section{Calculation and results}
\label{sec:calcres}

We have implemented the factorisation scheme transformation kernels,
the \msbar splitting functions,
and the required commutation integrals,
in a \Mathematica \cite{Mathematica} package, \texttt{DGLAPFS}.%
\footnote{\texttt{DGLAPFS} is available from the authors on request;
a future public release is anticipated.}
Internally, \texttt{DGLAPFS} makes
use of the \MT package \cite{Hoschele:2013pvt}
for \Mathematica
to perform Mellin convolutions,
and the \HPL package \cite{Maitre:2005uu,Maitre:2007kp} to manipulate and simplify
harmonic polylogarithms.

The transformation kernels from
Tables 1--4 of \cite{Delorme:2025teo}
are implemented directly in the form of \cref{eq:tableDecomp}.
The NLO \msbar splitting functions have been taken from
\cite{Ellis:1996mzs}
and cross-checked against those based on \cite{Moch:2004pa,Vogt:2004mw}
included within the \MT package, after substituting the colour-algebra factors.

We have calculated the NLO splitting functions
in all the schemes considered in \cite{Delorme:2025teo}.
For each scheme, we test that our calculated splitting functions
satisfy the appropriate sum rules, outlined in \cref{sec:sumrules}.
This is done numerically for the coefficient of each combination of colour-factors,
with a tolerance of $10^{-70}$.
We have further verified that the commutators we obtain from \cref{eq:PFS2ab}
for the \dis scheme match those given in Appendix C of \cite{Forte:2000wh}.

Here we present the schemes of greatest interest for phenomenological applications: \krk and \phys.
Additionally, we introduce the parametrised `generalised' scheme (\gen) allowing generalisation beyond the named schemes. 
Further, in \cref{sec:appMSbarDIS} we provide the corresponding results for the \dis scheme in the same conventions. We do so because to our knowledge, despite its historical significance, closed-form analytic expressions for splitting functions in the \dis scheme are not otherwise available in the literature.

We present our results decomposed by colour factors,
in terms of plus-distributions $\mathcal{D}_k$,
delta-functions, powers of logarithms, and the dilogarithm
\begin{align}
    \li2 (z) = - \int_0^z \frac{1}{t} \log (1-t) \, \dd t = \sum_{k=1}^\infty \frac{z^k}{k^2}.
\end{align}
As in \cite{Delorme:2025teo} we use plus-distribution identities to
ensure the coefficients of plus-distributions are numbers,
not functions of $x$.
Note that $x$ here is the argument of the splitting functions,
not Bjorken $x_B$.

We have checked that the expressions
as presented below yield the same result upon convolution with a test function as a
separate, entirely-numerical implementation of the
transformation convolutions,
again tested colour-factor by colour-factor.

\begin{figure*}[p]
\centering
\begin{subfigure}[t]{0.4\textwidth}
    \includegraphics[width=\textwidth]{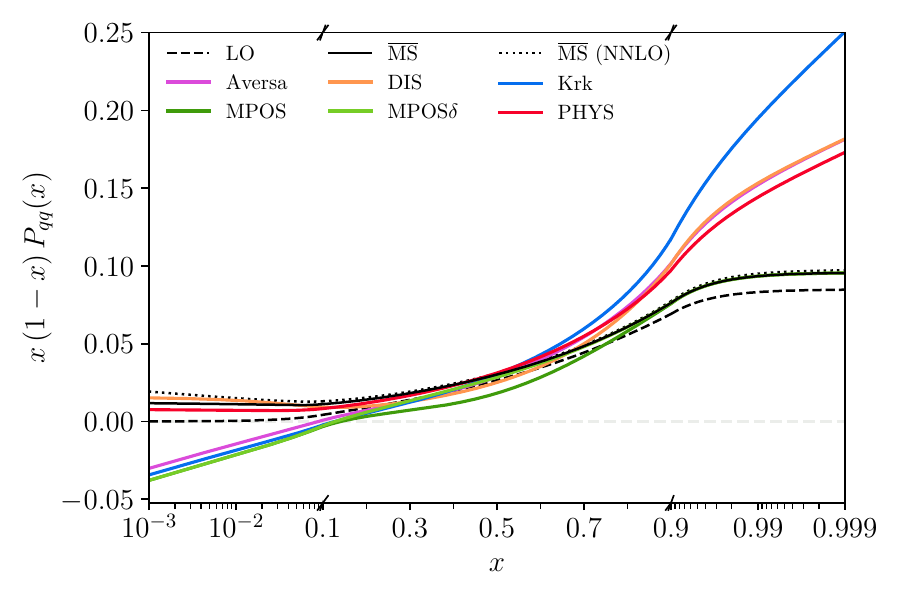}
\caption{$P_{qq}^{\rFS}$\label{fig:xspacePqq}}
\end{subfigure}
\begin{subfigure}[t]{0.4\textwidth}
    \includegraphics[width=\textwidth]{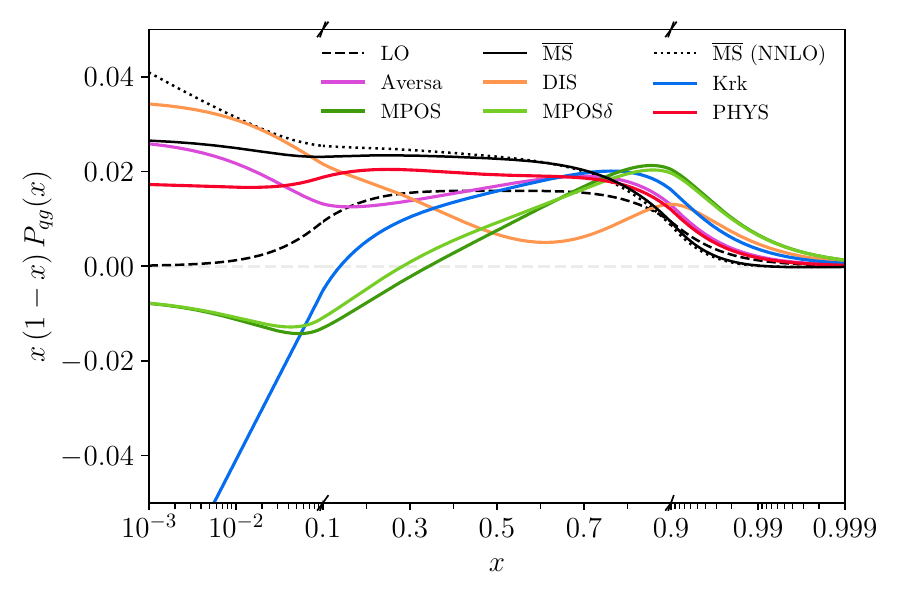}
\caption{$P_{qg}^{\rFS}$\label{fig:xspacePqg}}
\end{subfigure}
\\
\begin{subfigure}[t]{0.4\textwidth}
    \includegraphics[width=\textwidth]{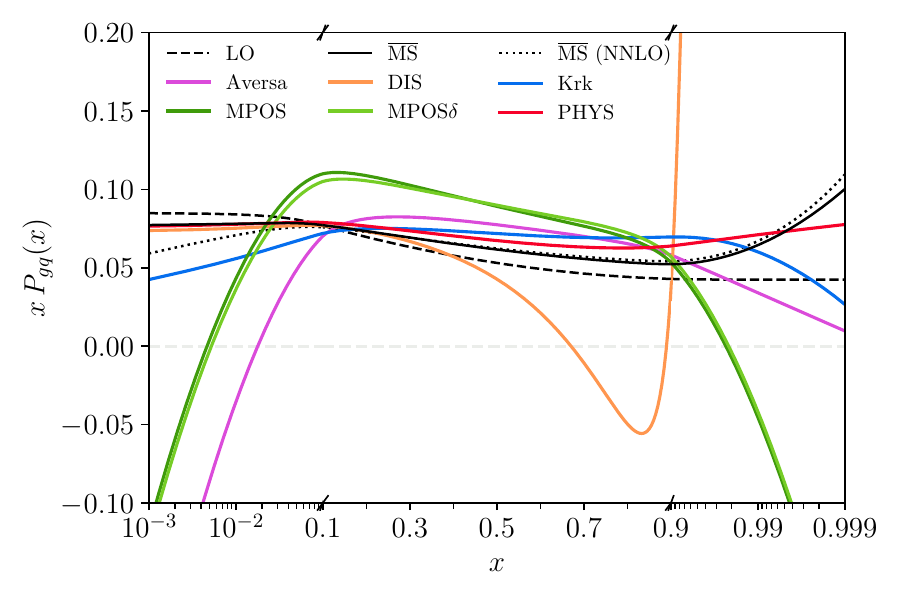}
\caption{$P_{gq}^{\rFS}$\label{fig:xspacePgq}}
\end{subfigure}
\begin{subfigure}[t]{0.4\textwidth}
    \includegraphics[width=\textwidth]{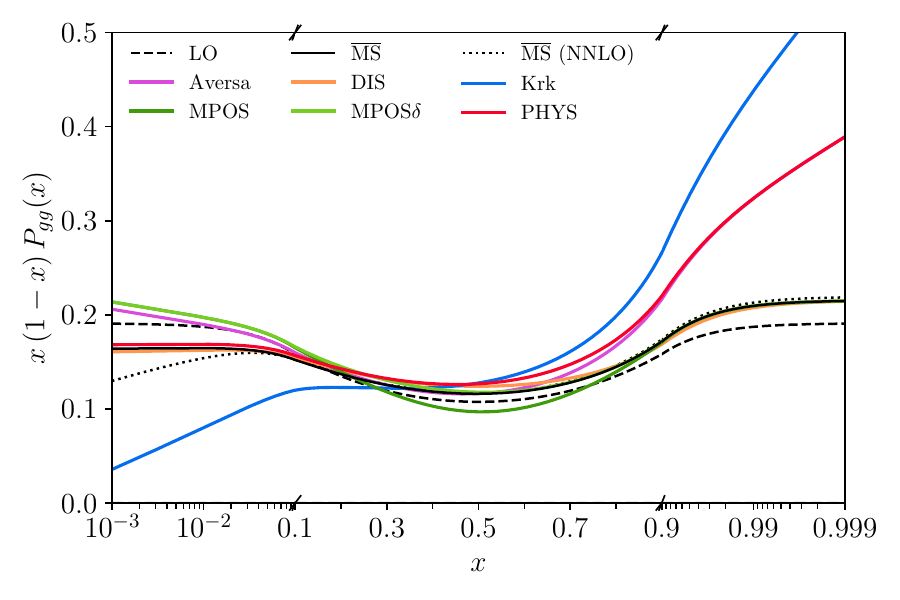}
\caption{$P_{gg}^{\rFS}$\label{fig:xspacePgg}}
\end{subfigure}
\caption{Splitting functions for the singlet sector
(calculated with $\alphas = 0.2$, $\nf = 4$), 
multiplied by $x (1-x)$ in the diagonal channels and for $qg$,
and by $x$ for $gq$ (due to its lesser degree of divergence as $x\to 1$).
}
\label{fig:xspaceP}
\end{figure*}

\begin{figure*}[p]
\centering
\begin{subfigure}[t]{0.4\textwidth}
    \includegraphics[width=\textwidth]{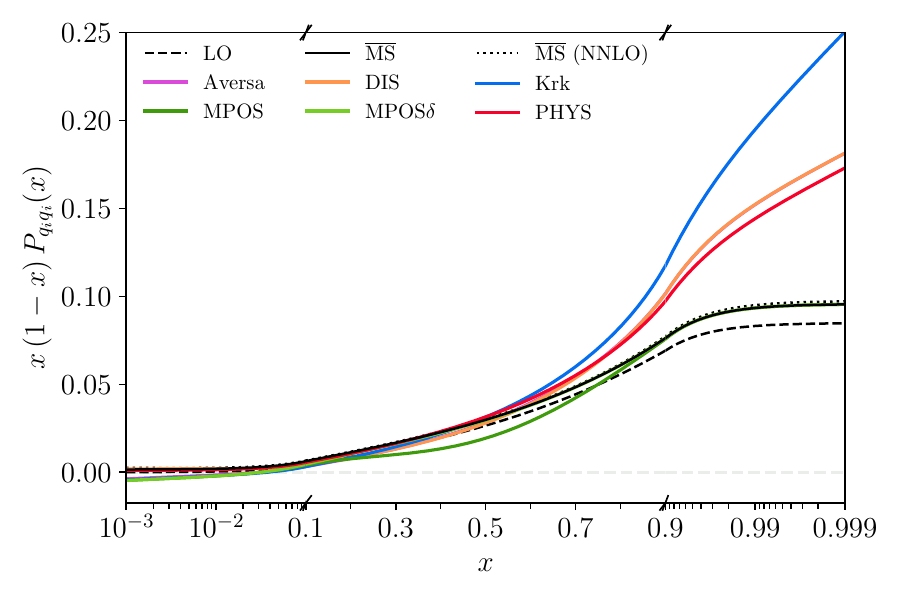}
\caption{$P_{q_iq_i}^{\rFS}$\label{fig:xspacePqiqi}}
\end{subfigure}
\begin{subfigure}[t]{0.4\textwidth}
    \includegraphics[width=\textwidth]{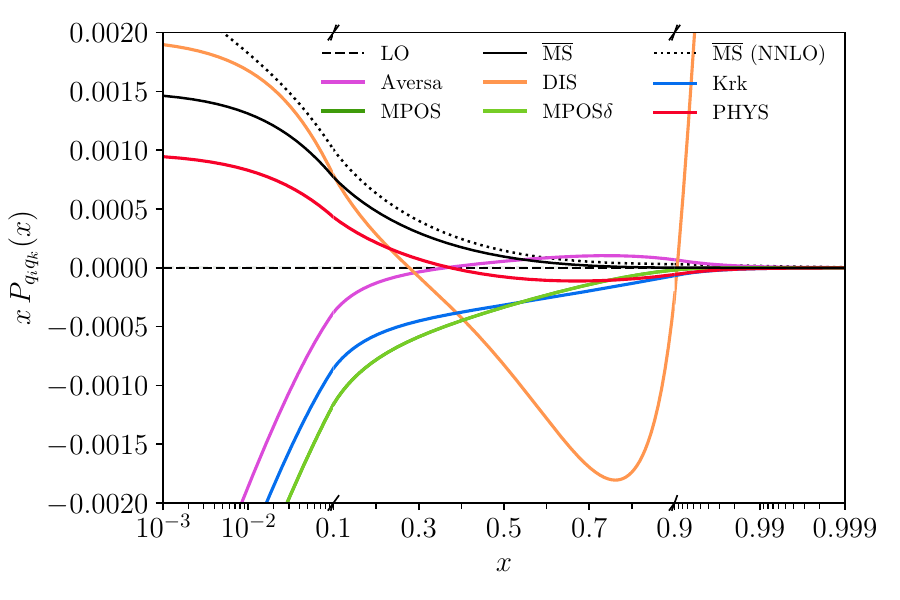}
\caption{$P_{q_iq_k}^{\rFS}$\label{fig:xspacePqiqk}}
\end{subfigure}
\\
\begin{subfigure}[t]{0.4\textwidth}
    \includegraphics[width=\textwidth]{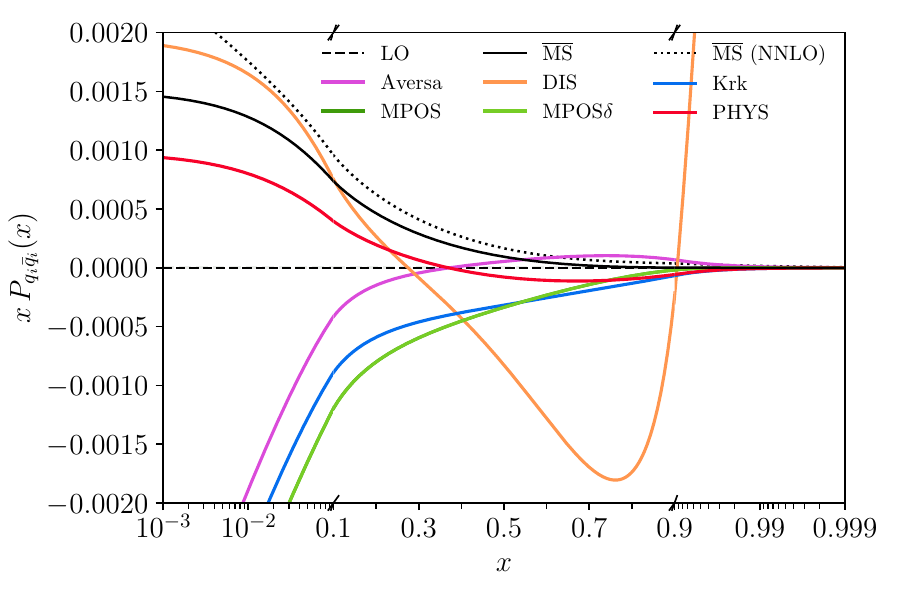}
\caption{$P_{q_i\overline{q}_i}^{\rFS}$\label{fig:xspacePqiqbi}}
\end{subfigure}
\begin{subfigure}[t]{0.4\textwidth}
    \includegraphics[width=\textwidth]{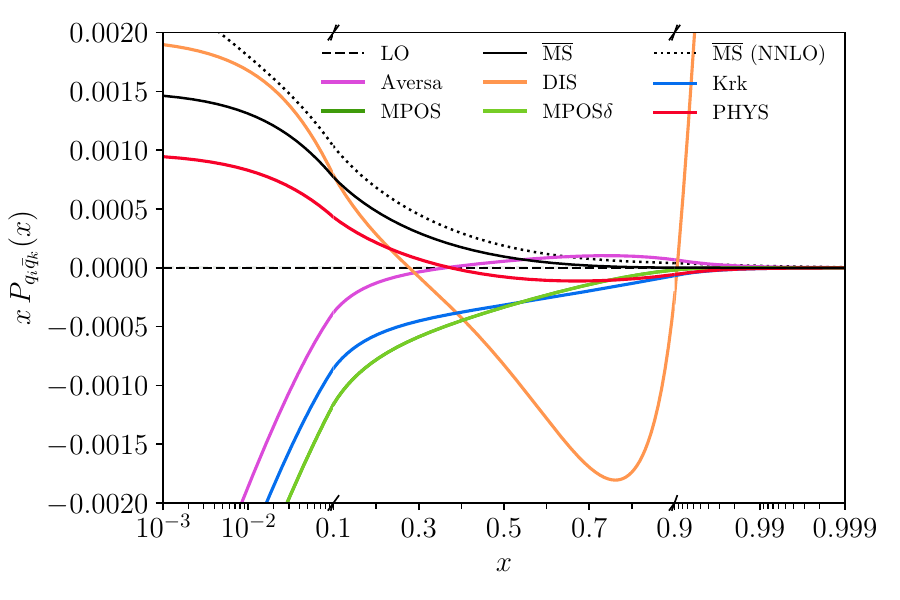}
\caption{$P_{q_i\overline{q}_k}^{\rFS}$\label{fig:xspacePqiqbk}}
\end{subfigure}
\caption{Splitting functions for the non-singlet sector (calculated with $\alphas = 0.2$, $\nf = 4$), multiplied by $x$ for the off-diagonal channels, and $x(1-x)$ for the diagonal channel.
Note that at NLO, 
$P_{q_iq_k}^{\rFS} \equiv
P_{q_i\overline{q}_k}^{\rFS}$.
Although $P_{q_i\overline{q}_i}^{\rFS} \neq P_{q_iq_k}^{\rFS}$, numerically they are very close.}
\label{fig:xspacePq}
\end{figure*}

The calculated splitting functions are plotted in \cref{fig:xspaceP,fig:xspacePq}, multiplied by $x(1-x)$ where necessary to tame the singularities at 0 and 1,
for $\alphas = 0.2$ and $\nf = 4$ (adopting the conventions established in \cite{Moch:2004pa,Vogt:2004mw}).%
\footnote{Note that the scale used for the $x$-axis in \cref{fig:xspaceP,fig:xspacePq} is symmetric around $x=0.5$, with a linear scale from 0.1 to 0.9, and appropriate logarithmic scales approaching both relevant limits, 0 and 1.
Functions proportional to $\log x$ therefore
appear linear as $x$ approaches 0; likewise,
functions proportional to $\log(1-x)$ appear linear as $x$ approaches 1.
Because of the $x(1-x)$ factor, these correspond to the leading small-$x$ logarithmic terms, and the threshold $\mathcal{D}_1$ plus-distribution terms, respectively.}
For reference we also include the
scheme-independent LO splitting functions,
and the \msbar splitting functions at NLO and NNLO.

It is immediately clear that the largest differences between the splitting functions emerge in the low- and high-$x$ limits,
where the splitting functions can diverge as
$(\log^k x)/x$ and $(\log^k (1-x))/(1-x)$ respectively.
This will be discussed in detail when analysing their asymptotic behaviour in Sec.~\ref{sec:asymptotics}.
In the interior of the unit interval, the boundedness of the perturbative transformation
leads to less dramatic differences between the schemes.

The singlet sector, shown in \cref{fig:xspaceP}, can be divided into 
diagonal and off-diagonal channels.
In the diagonal channels, the schemes divide into three emergent groups
with differing degrees of divergence as $x \to 1$, and two groups as $x \to 0$.
In the off-diagonal channels, the $x \to 1$ divergence is generally suppressed.
In the quark sector, the single diagonal channel has the
$x\to 1$ limiting behaviour observed in \cref{fig:xspacePqq}
and the $x \to 0$ behaviour arises from the off-diagonal flavour channels
(which are zero in the $x\to 1$ limit).
These features will be visualised in Mellin space in
\cref{sec:anomalousdims} and are the subject of the asymptotic analysis in \cref{sec:asymptotics}.

\subsection{Krk scheme}
\label{sec:results_krk}
The \krk scheme \cite{Jadach:2015mza,Jadach:2016qti,Jadach:2016acv} uses for its factorisation scheme transformation the collinear convolution terms arising from the integral of Catani--Seymour subtraction dipoles \cite{Catani:1996vz}
over the unresolved phase space
for colour-singlet final states. By absorbing these collinear counterterms into the PDFs rather than including them in the hard-process, matched NLO-plus-parton-shower calculations can be implemented using only multiplicative weights (the \krknlo method \cite{Jadach:2015mza,Jadach:2016qti,Sarmah:2024hdk,Sarmah:2025vnb}).
NLO splitting functions in the \krk scheme are given in
\cref{eq:PKrkqiqi,eq:PKrkqiqk,eq:PKrkqiqbi,eq:PKrkqig,eq:PKrkgq,eq:PKrkgg}.

\input{splitfns/PKrk_formatted.tex}

\subsection{PHYS scheme}
\label{sec:results_phys}
The \phys scheme was introduced in \cite{Oliveira:2013aug} to remove finite ${\varepsilon}/{\varepsilon}$ contributions,
of IR origin which emerge in the \msbar scheme from long-distance interactions between massless QCD partons.
NLO splitting functions in the \phys scheme are given in
\cref{eq:PPHYSqiqi,eq:PPHYSqiqk,eq:PPHYSqiqbi,eq:PPHYSqig,eq:PPHYSgq,eq:PPHYSgg}.

\input{splitfns/PPHYS_formatted.tex}

\subsection{Generalised scheme}
\label{sec:results_gen}

Following \cite{Delorme:2025teo} we parametrise possible 
scheme transformations in order to generalise beyond the named schemes
considered there:
\begin{align}
\label{eq:genschemeKqq}
    \rK_{qq}^{\msbar\to\gen} (z) & {} =
        \cf \biggl[
            a_{qq} \biggl( 2 \mathcal{D}_1 - (1+z) \log(1-z) \biggr) 
            - b_{qq} \mathcal{D}_0
            - c_{qq} \pqq \log z
            - \Delta_{qq} \delta(1-z)
        \biggr],
    \\ \label{eq:genschemeKqg}
    \rK_{qg}^{\msbar\to\gen} (z) & {} = 
        \tr \biggl[
            a_{qg} \pqg \log(1-z) - c_{qg} \pqg \log z
        \biggr],
    \\ \label{eq:genschemeKgq}
    \rK_{gq}^{\msbar\to\gen} (z) & {} =
        \cf \biggl[
            a_{gq} \pgq \log(1-z) - c_{gq} \pgq \log z
        \biggr],
    \\ \label{eq:genschemeKgg}
    \rK_{gg}^{\msbar\to\gen} (z) & {} =
        \ca \biggl[
            2 a_{gg} \biggl( \mathcal{D}_1 + \left(\frac{1}{z} - 2 + z(1-z)\right)
             \log(1-z) \biggr)
             - b_{gg} \mathcal{D}_0
            - 2 c_{gg} \pgg \log z
            - \Delta_{gg} \delta(1-z)
        \biggr],
\end{align}
where the values of the coefficients for specific named schemes are provided in \cref{tab:FScoeffs}.
The terms which do not follow the ansatz of \cref{eq:genschemeKqq,eq:genschemeKqg,eq:genschemeKgq,eq:genschemeKgg} are indicated in the Table by `---'.

As discussed in \cite{Delorme:2025teo}, the \dis\ and \aversa\ schemes cannot be fully represented in such a general scheme,
since $\rK_{ga}^{\msbar\to\dis}$ mixes splitting-functions between channels (due to the imposition of the
moment constraint extending the momentum sum rule to all $N$), and
$\rK_{gg}^{\msbar\to\aversa}$ uses only a
single (rational) term from $\pgg$ in its coefficient of $\log z$.

This parametrisation neglects the $P(z)$ contribution in \cref{eq:tableDecomp}, consistently setting it to 0 across
all schemes and channels.
Since $z \in (0,1]$, polynomial terms $z^k$ for $k\geqslant 0$ are bounded by the sum of their coefficients,
and give suppressed contributions $1/(N+k)$ in Mellin space;
pole terms $1/z$ are in principle possible%
\footnote{Of the schemes considered in \cite{Delorme:2025teo},
they arise only in the $gq$ kernels in the \pos family of schemes, from $\pgq$.}
but would change the Mellin-space pole structure of all coefficient functions in that scheme, and so are excluded.

Within this parametrisation,
modulo the restriction of the $P(z)$ terms,
the schemes discussed in \cite{Delorme:2025teo} may be
considered special cases of the parametrised form in \cref{eq:genschemeKqq,eq:genschemeKqg,eq:genschemeKgq,eq:genschemeKgg}
with $a_{qq},a_{gg} \in \{ 0, 1, 2\}$,
$a_{qg}, a_{gq} \in \{1,2\}$,
$b_{qq} \in \{0,\frac{3}{2}\}$,
$b_{gg} = 0$,%
\footnote{In \cite{Delorme:2025teo} this parameter was neglected; such a term is not present in any of the concrete schemes considered. We include it here for consistency with the $qq$-case,
to streamline the discussion in \cref{sec:asymptotics},
but exclude it from the dimension counting.}
$c_{qq}, c_{qg}, c_{gq}, c_{gg} \in \{0, 1\}$,
and $\Delta_{qq}, \Delta_{gg}$ often fixed by the imposition of
the momentum sum rule as in \cite{Delorme:2025teo}, but here left unconstrained.

\begin{table*}[tbp]
	\begin{center}
		\begin{tabular}{ |c|c|c|c|c|c|c|c|c|c|c| } 
        \hline
            Scheme
			& $a_{qq}$
			& $a_{gg}$
			& $a_{qg}$
			& $a_{gq}$
			& $b_{qq}$
			& $b_{gg}$
			& $c_{qq}$
			& $c_{gg}$
			& $c_{qg}$
			& $c_{gq}$
			\\
			\hline
            \msbar
            & 0
            & 0
            & 0
            & 0
            & 0
            & 0
            & 0
            & 0
            & 0
            & 0
			\\
            \aversa
            & 1
            & 1
            & 1
            & 1
            & $\frac{3}{2}$
            & 0
            & 1
            & ---
            & 1
            & 1
            \\
            \dis
            & 1
            & 0
            & 1
            & ---
            & $\frac{3}{2}$
            & 0
            & 1
            & ---
            & 1
            & ---
            \\
            \krk
            & 2
            & 2
            & 2
            & 2
            & 0
            & 0
            & 1
            & 1
            & 1
            & 1
            \\
            \mpos($\delta$)
            & 0
            & 0
            & 2
            & 2
            & 0
            & 0
            & 0
            & 0
            & 1
            & 1
            \\ 
            \phys
            & 1
            & 1
            & 1
            & 1
            & 0
            & 0
            & 0
            & 0
            & 0
            & 0
            \\ \hline
        \end{tabular}
    \end{center}
    \caption{Fully-defined factorisation schemes from \cite{Delorme:2025teo},
    expressed in the notation of \cref{eq:genschemeKqq,eq:genschemeKqg,eq:genschemeKgq,eq:genschemeKgg}.
    The rational piece $P(z)$ has been omitted, as has the endpoint
    contribution $\Delta_{aa}$ (typically constrained to be a function of the other parameters, to impose momentum conservation).
    `---' indicates that the kernel does not fit the corresponding ansatz and so cannot be expressed by any choice of coefficient.
	\label{tab:FScoeffs}}
\end{table*}

This gives a physically-motivated nine-parameter%
\footnote{Omitting $b_{gg}$ and treating $\Delta_{qq}, \Delta_{gg}$ as dependent parameters as discussed in \cref{sec:sumrules}.}
subspace of the space of possible factorisation schemes,
and a finite subset within them of parameters known to arise in named schemes.
Save for the \dis scheme, the fully-defined schemes considered in \cite{Delorme:2025teo} satisfy the further constraints
\begin{equation}
    a_{qq} = a_{gg}, \qquad a_{qg} = a_{gq}
\end{equation}
and the \krk, \mpos($\delta$) and \phys schemes satisfy
\begin{equation}
    c_{qq} = c_{gg}, \qquad c_{qg} = c_{gq}.
\end{equation}
These constraints further reduce the dimensionality of the parameter space,
and will be seen to simplify the resulting splitting functions by eliminating
certain contributions related to threshold logarithms ($a$) and high-energy logarithms ($c$).

The \msbar splitting functions may be recovered by setting the scheme-transformation
coefficients to zero.
As remarked upon in \cref{sec:facSchemes}, the scheme-transformation of \cref{eq:PFS2ab}
is valid for both spacelike evolution, as considered here, and timelike evolution as required for
fragmentation functions. These \gen-scheme splitting functions can therefore be adapted for timelike evolution
by adding the usual spacelike-to-timelike conversion functions used for the \msbar scheme \cite{Curci:1980uw,Furmanski:1980cm}.%
\footnote{
To the extent that the \gen-scheme ansatz of \cref{eq:genschemeKqq,eq:genschemeKqg,eq:genschemeKgq,eq:genschemeKgg} is also well-motivated for timelike evolution, this immediately enables a study of factorisation-scheme dependence also for fragmentation function evolution.}
In the following we present the spacelike splitting functions in the \gen scheme:

\input{splitfns/PGEN_formatted.tex}

\subsection{Sum rules}
\label{sec:sumrules}

Both the familiar {\em momentum sum rule}
\begin{align}
	\label{eq:momsumrule_PDFs_MSbar}
	\sum_a \langle x_a \rangle^\msbar (\mu) = \sum_a \int_0^1 x \, f_{a}^{\msbar} (x, \mu) \; \dd x = 1 ,
\end{align}
and the {\em valence-quark number sum rule}, for the proton,
\begin{align}
	\label{eq:numbersumrule_PDFs_MSbar}
	N_q^{\msbar}
    \colonequals
    \int_0^1 \dd x \, \left[ f_{q}^{\msbar}(x;\muf) - f_{\bar{q}}^{\msbar}(x;\muf) \right] = 
    \begin{cases}
        2, & \quad q=u \\
        1, & \quad q=d \\
        0, & \quad \text{else}
    \end{cases}
\end{align}
(assuming there are no asymmetric intrinsic contributions, e.g.\ for charm),
are scheme-dependent. They can be proven to hold in the \msbar scheme
(see, for example, \cite{Collins:2011zzd}),
but in general are modified by factorisation-scheme transformations, obtaining a correction starting at NLO,
\begin{align}
	\label{eq:momsumrule_PDFs}
    \sum_a \langle x_a \rangle^{\rFS} (\mu)  & {} %
    = 1 
    + \frac{\alphas(\mu)}{2\pi}
    \sum_b
    \langle x_b\rangle^{\msbar} (\mu) 
    \sum_a \int_0^1 \;  z \, \rK_{ab}^{\msbar\to\rFS}(z) \; \dd z,
\\ \label{eq:numbersumrule_PDFs_FS}
 N_q^{\rFS} (\mu)  & {} = 
   N_q^{\msbar} \left\{ 1 + \frac{\alphas(\mu)}{2\pi} \,
  \int_0^1 \dd z \; \rK^{\msbar \to \rFS}_{qq}(z) \right\}.
\end{align}

In order to compare factorisation schemes on a like-for-like basis,
in \cite{Delorme:2025teo} we consistently imposed the \msbar momentum sum rule \cite{Collins:1981uw} at the level of the transformation kernels (this is also done in the original formulation of most of the considered schemes),
\begin{align}
	\label{eq:momsumrule_kernels}
    \sum_a \int_0^1 \; z \, \rK_{ab}^{\msbar\to\rFS}(z) \; \dd z = 0
\end{align}
for all flavours $b$. This is obtained using the virtual-type endpoint coefficient $\Delta$
for the flavour-diagonal kernels where necessary,
\begin{align}
	\label{eq:momsumrule_deltas}
    \rK_{bb}^{\prime\msbar\to\rFS}(z)
    & = \rK_{bb}^{\msbar\to\rFS}(z) 
    - 
    \delta(1-z) \sum_a \int_0^1 \; z' \, \rK_{ab}^{\msbar\to\rFS}(z') \; \dd z'.
\end{align}
For the generalised \gen scheme defined in \cref{sec:results_gen}, this implies%
\footnote{These expressions can be reconciled with the values for the named schemes given in \cite{Delorme:2025teo} by combining them with the
integrals of the missing $P(z)$ terms.}
\begin{align}
    \label{eq:GENDeltaqq}
    \Delta_{qq} & {} = 
    \frac{121}{36} a_{qq} + b_{qq} + \left( \frac{\pi^2}{3} - \frac{85}{36} \right) c_{qq} - \frac{10}{9} a_{gq} + \frac{29}{18} c_{gq} ,
    \\
    \Delta_{gg} & {} = 
    \frac{203}{72} a_{gg} + b_{gg} + \left( \frac{\pi^2}{3} - \frac{65}{72} \right) c_{gg} + \frac{\tr\nf}{36 \ca} \left( - 41 a_{qg} + 11 c_{qg} \right).
\end{align}

The \msbar valence-quark number sum rule cannot necessarily be imposed simultaneously
as an endpoint contribution;%
\footnote{
Modifications to the transformation kernel in the bulk $z \in (0,1)$ can in principle be constructed to enable this, e.g.\ using functions odd about $z=\frac{1}{2}$ to impose momentum conservation.}
unless $\rK_{qq}$ integrates to 0
(for instance when $\rK_{qq}$ is given by a plus distribution, as for the \dis{} scheme).
When this is not the case,
$N_q^{\rFS}$ is modified by terms of order $\order{\alphas}$ relative to $N_{q}^\msbar$, and hence becomes scale-dependent, as shown in \cref{eq:numbersumrule_PDFs_FS}.
Due to the running of $\alphas$, this implies that the \msbar result is recovered in the high-energy limit, in any scheme.

At the level of the splitting functions, 
taking the derivative of \cref{eq:momsumrule_PDFs_MSbar}
implies
\begin{align}
	\label{eq:momsumrule_PMSbar}
    \sum_a \int_0^1 \; z \, \rP_{ab}^{\msbar}(z) \; \dd z = 0
\end{align}
for all $b$, in order for evolution to preserve the sum rules independently of the
specific PDFs used.
Adapted to a general scheme using \cref{eq:DGLAPFStrans},
defining
\begin{align}
R_b = \sum_a \int_0^1 \; z \, \rK_{ab}^{\msbar\to\rFS}(z) \; \dd z 
\end{align}
for the residual ($R_b=0$ for all $b$ if \cref{eq:momsumrule_kernels} is imposed),
at NLO this becomes
\begin{align}
	\label{eq:momsumrule_P}
    \sum_a \int_0^1 \; z \, \rP_{ab}^{\rFS(2)}(z) \; \dd z 
    & {} =
    \frac{\beta^{(2)}}{2\pi}
    R_b
    + 
    \sum_c R_c \,
    \int_0^1 \; z \, \rP_{cb}^{(1)}(z) \; \dd z .
\end{align}

As a consequence, if the \msbar momentum-conservation sum rule is imposed on the transformation kernels, the splitting functions also satisfy the \msbar momentum-conservation sum rule.

The valence quark-number sum rule corresponds instead to the first Mellin moment
of the non-singlet, $f_q - f_{\overline{q}}$, and so
the contribution from the commutator vanishes.
The evolution equation therefore proceeds as expected from \cref{eq:numbersumrule_PDFs_MSbar},
\begin{align}
    \label{eq:numbersumrule_Ps_FS}
    \int_0^1 \left( \rP^{\rFS(2)}_{q_iq_i}(z) - \rP^{\rFS(2)}_{q_i \overline{q}_i}(z) \right) \, \dd z = \frac{\beta^{(2)}}{2\pi}
    \int_0^1 \rK^{\msbar \to \rFS}_{qq}(z) \; \dd z.
\end{align}
The values taken by this sum rule in the various schemes are
given in \cref{tab:intKqq}.

\begin{table*}[tbp]
\begin{center}
\begin{tabular}{|c|c|c|c|c|c|c|c|c|} 
\hline
$ $
& \msbar
& \aversa
& \dis
& \krk
& \mpos
& \mposd
& \phys
& \gen
\\ \hline
$ \displaystyle  \frac{2\pi}{\cf \beta^{(2)}} \int_0^1 \left( \rP^{\rFS(2)}_{q_iq_i} - \rP^{\rFS(2)}_{q_i \overline{q}_i} \right) \; \dd z $
& $0$
& $0$
& $0$
& $ \displaystyle -\frac{3}{2}$
& $ \displaystyle \frac{35}{9}$
& $ \displaystyle \frac{35}{18}$
& $ \displaystyle - \frac{1}{2}$
& $\frac{7}{4} a_{qq} + \left( \frac{\pi^2}{3} - \frac{5}{4} \right) c_{qq} - \Delta_{qq}$
\\ \hline
\end{tabular}
\end{center}
\caption{Integral of the $\rK_{qq}^{\rFS}(z)$ kernel entering the valence-quark number sum rule in \cref{eq:numbersumrule_Ps_FS}.
A value of 0 corresponds to the $\msbar$ sum rule, unmodified.}
\label{tab:intKqq}
\end{table*}

\begin{figure*}[bp]
\centering
\begin{subfigure}[t]{0.46\textwidth}
    \includegraphics[width=\textwidth]{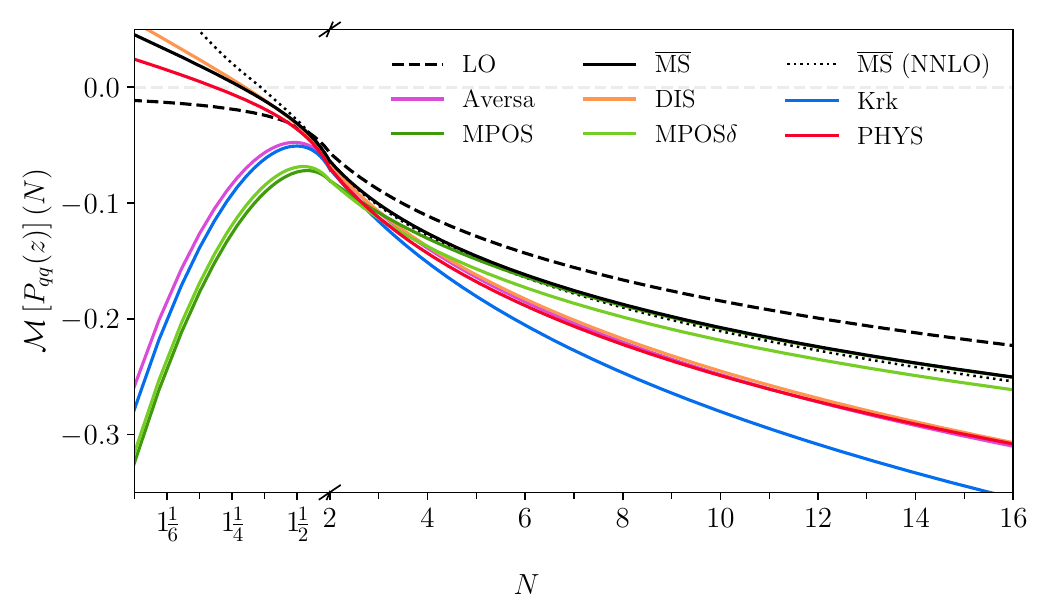}
\caption{$P_{qq}^{\rFS}$\label{fig:NspacePqq}}
\end{subfigure}
\begin{subfigure}[t]{0.46\textwidth}
    \includegraphics[width=\textwidth]{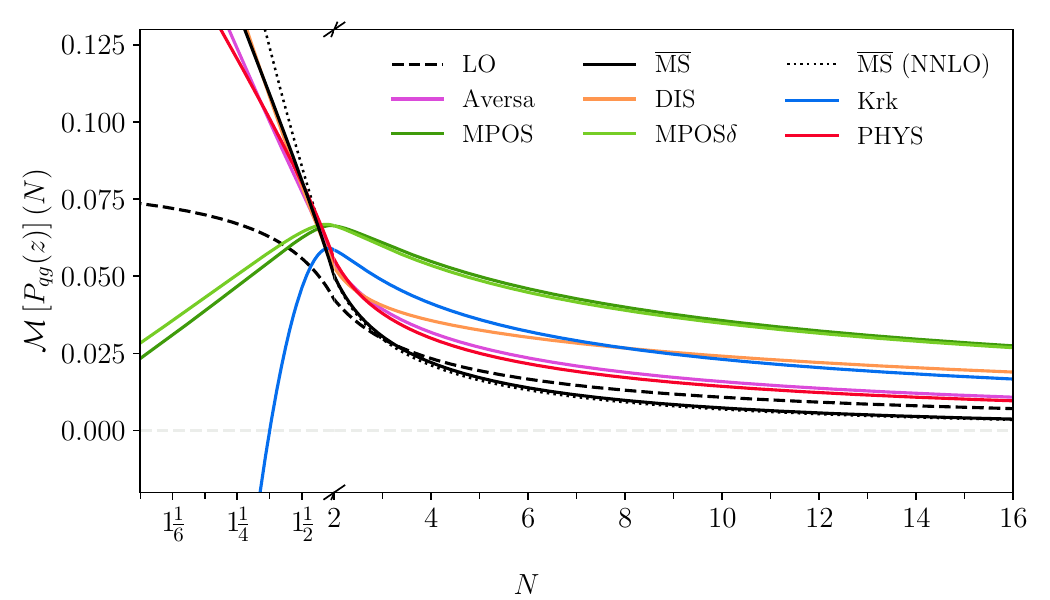}
\caption{$P_{qg}^{\rFS}$\label{fig:NspacePqg}}
\end{subfigure}
\\
\begin{subfigure}[t]{0.46\textwidth}
    \includegraphics[width=\textwidth]{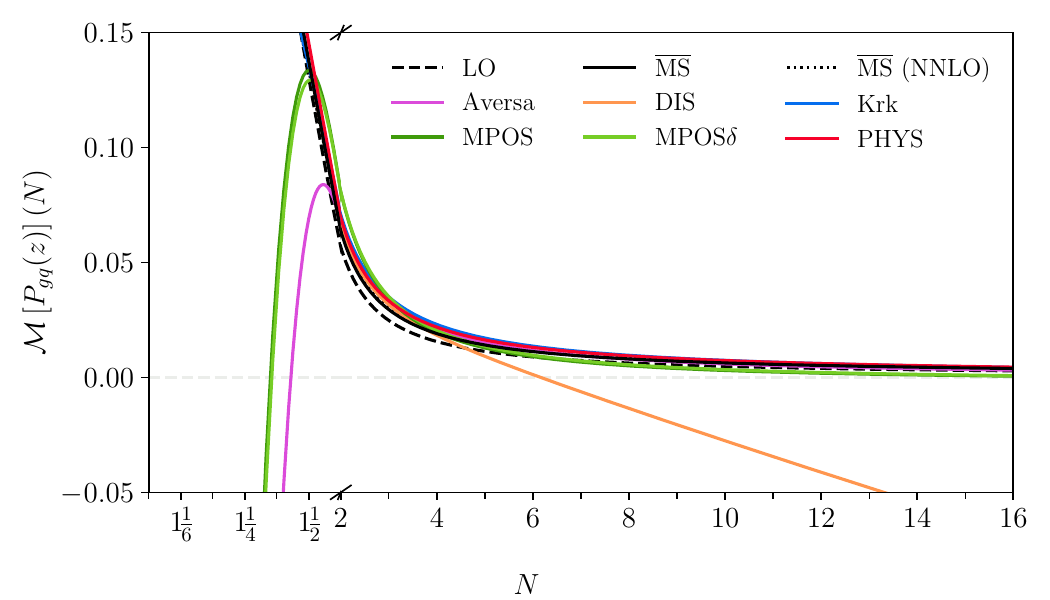}
\caption{$P_{gq}^{\rFS}$\label{fig:NspacePgq}}
\end{subfigure}
\begin{subfigure}[t]{0.46\textwidth}
    \includegraphics[width=\textwidth]{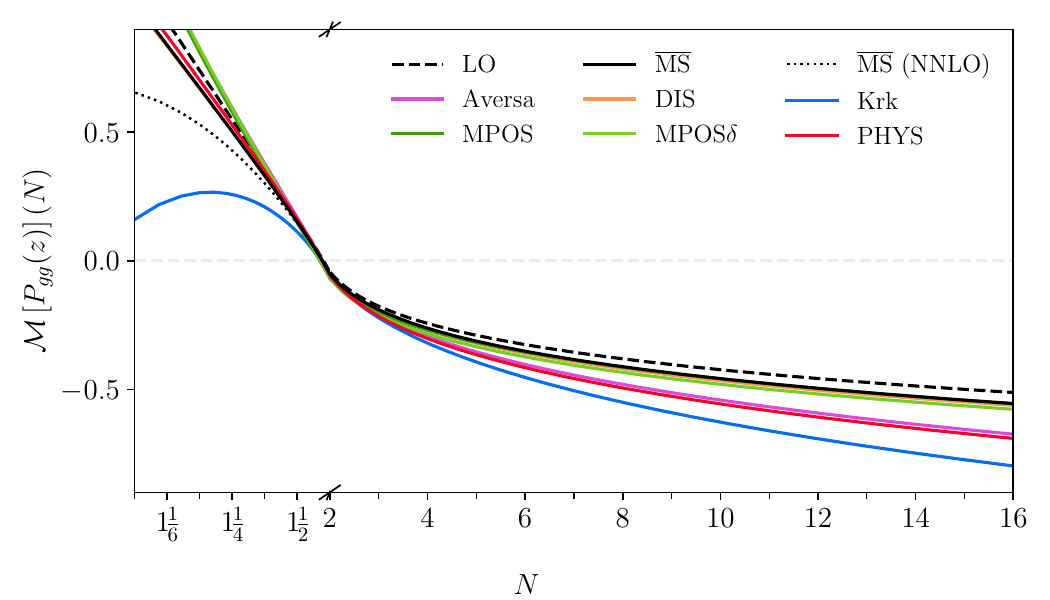}
\caption{$P_{gg}^{\rFS}$\label{fig:NspacePgg}}
\end{subfigure}
\caption{Anomalous dimensions for the singlet sector,
with reciprocal-scaled axes below $N=2$
(calculated with $\alphas = 0.2$, $\nf = 4$).
}
\label{fig:NspaceP}
\end{figure*}

\begin{figure*}[p]
\centering
\begin{subfigure}[t]{0.46\textwidth}
    \includegraphics[width=\textwidth]{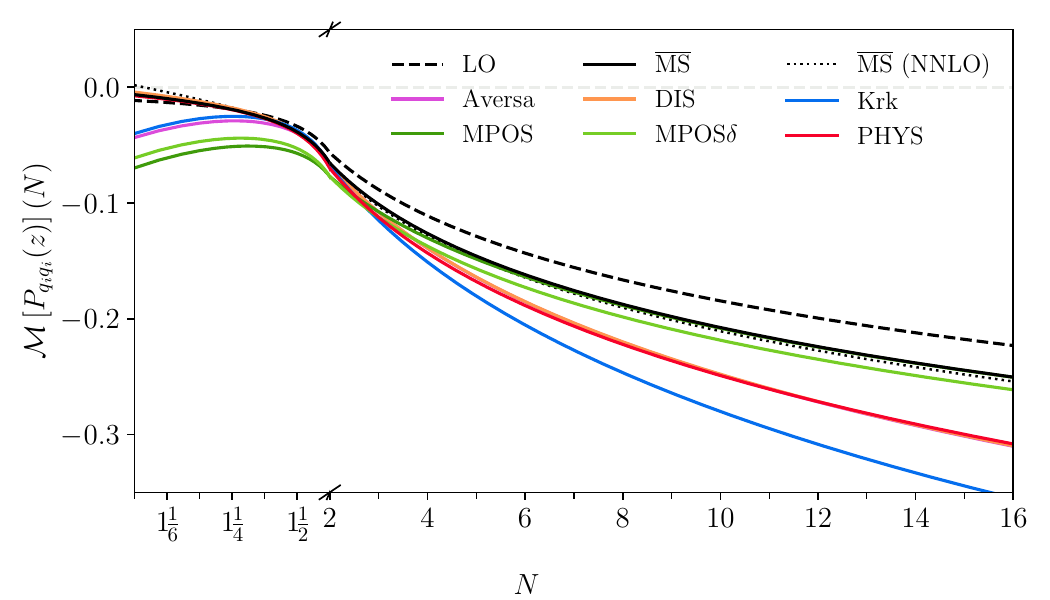}
\caption{$\gamma_{q_iq_i}^{\rFS} = \mathcal{M}[P_{q_iq_i}^{\rFS}]$\label{fig:NspacePqiqi}}
\end{subfigure}
\begin{subfigure}[t]{0.46\textwidth}
    \includegraphics[width=\textwidth]{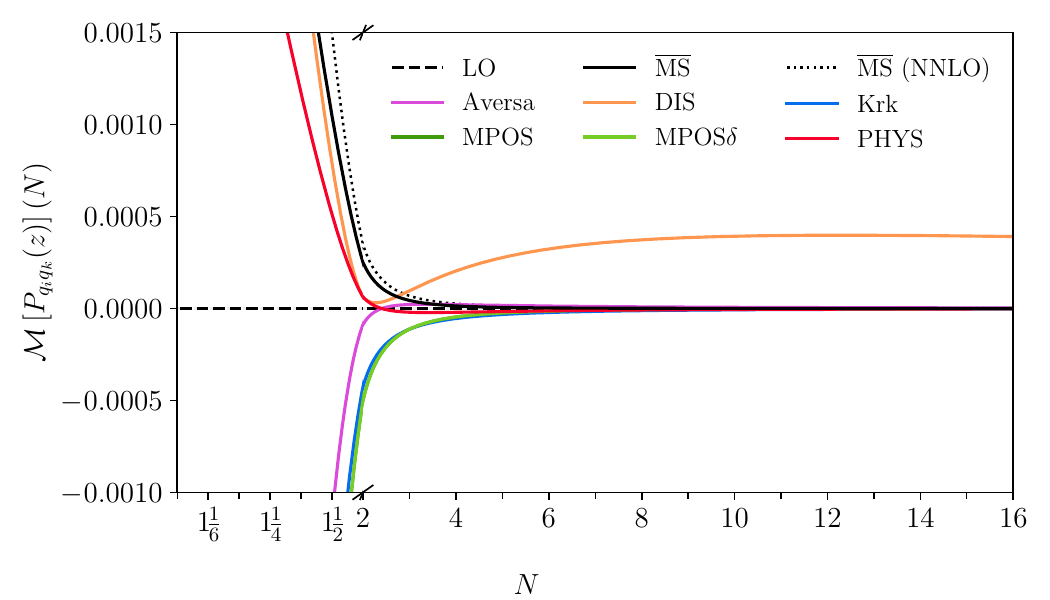}
\caption{$\gamma_{q_iq_k}^{\rFS} = \mathcal{M}[P_{q_iq_k}^{\rFS}]$\label{fig:NspacePqiqk}}
\end{subfigure}
\\
\begin{subfigure}[t]{0.46\textwidth}
    \includegraphics[width=\textwidth]{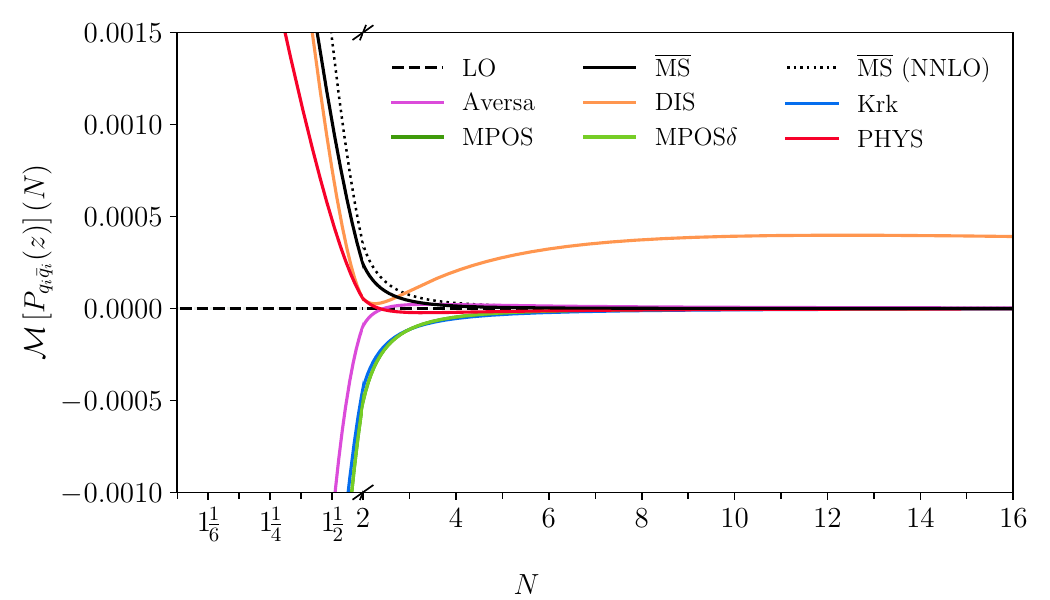}
\caption{$\gamma_{q_i\overline{q}_i}^{\rFS} = \mathcal{M}[P_{q_i\overline{q}_i}^{\rFS}]$\label{fig:NspacePqiqbi}}
\end{subfigure}
\begin{subfigure}[t]{0.46\textwidth}
    \includegraphics[width=\textwidth]{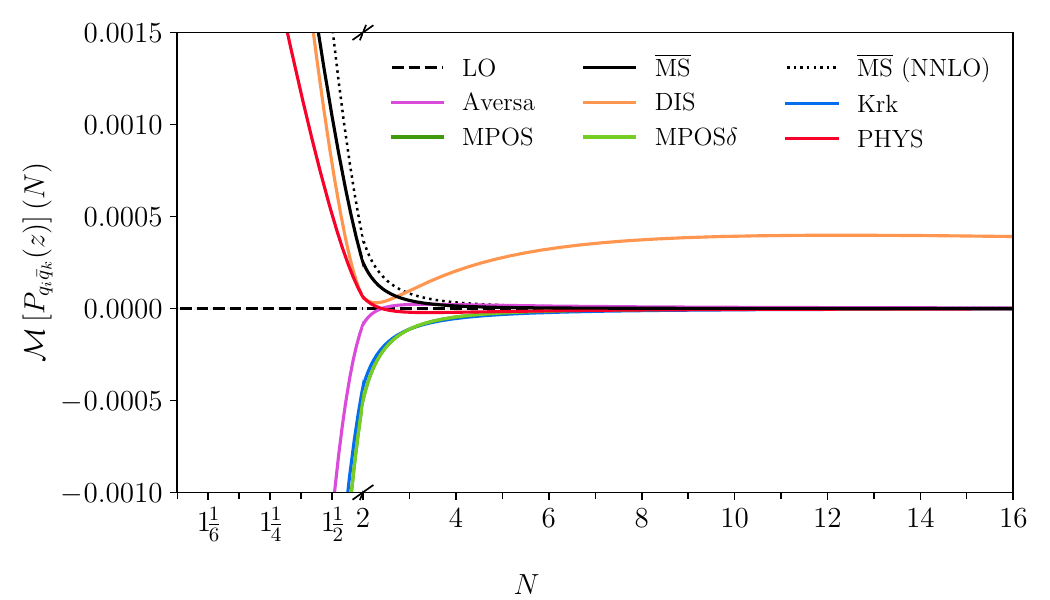}
\caption{$\gamma_{q_i\overline{q}_k}^{\rFS} = \mathcal{M}[P_{q_i\overline{q}_k}^{\rFS}]$\label{fig:NspacePqiqbk}}
\end{subfigure}
\caption{Anomalous dimensions for the non-singlet sector, with reciprocal-scaled axes below $N=2$  (calculated with $\alphas = 0.2$, $\nf = 4$).
Note that at NLO, 
$P_{q_iq_k}^{\rFS} \equiv
P_{q_i\overline{q}_k}^{\rFS}$.
Although $P_{q_i\overline{q}_i}^{\rFS} \neq P_{q_iq_k}^{\rFS}$, numerically they are very close, especially in Mellin space as plotted here.
}
\label{fig:NspacePq}
\end{figure*}

\section{Anomalous dimensions}
\label{sec:anomalousdims}

In \cref{fig:NspaceP,fig:NspacePq} we show the anomalous dimensions for the splitting
functions in each scheme considered in \cite{Delorme:2025teo},
\begin{align}
    \gamma_{ab}^{\rFS(k)}(N)
    = 
    \mathcal{M}[\rP^{\rFS(k)}_{ab}](N)
    =
    \int_0^1 z^{N-1} \, \rP^{\rFS(k)}_{ab}(z) \; \dd z.
\end{align}
We plot these at LO and at NLO in each scheme, restoring the associated factors of $\alphas$.
Following \cite{Moch:2004pa,Vogt:2004mw} we set $\alphas = 0.2$ and $\nf = 4$.

Immediately it is clear that in the singlet sector,
shown in \cref{fig:NspaceP}, the \msbar scheme is atypical
among the schemes in its consistent similarity to the leading-order
splitting functions indicating small perturbative corrections. 
As will be discussed analytically in \cref{sec:asymptotics},
this is largely due to the
absence of higher-order plus-distributions (beyond $\mathcal{D}_0$) to all orders in the \msbar scheme,
leading to perturbative corrections that are asymptotically
constant at large-$N$.
In a general factorisation scheme, 
the transformation kernels may reasonably contain plus-distributions up to the
order that would be encountered in a perturbative calculation of the same
order (at NLO, $\mathcal{D}_1$),
and hence these terms can also be present in the splitting functions.
However, splitting functions in the \msbar scheme have $\mathcal{D}_0$ as their most singular term (equivalently, $\log N$ in Mellin space), to all perturbative orders.

For the same reason, the differing large-$N$ behaviour between the schemes in the diagonal channels is due to the $a_{qq}$ and $a_{gg}$ parameters for the $qq$ and $gg$ splitting functions respectively.%
\footnote{This is the counterpart to the differences between
the coefficient functions observed in \cite{Delorme:2025teo}.}
This leads to a visible divergence between the schemes
with $a_{aa}=0$ (\msbar, \mpos($\delta$)),
$a_{aa}=1$ (\aversa, \dis, \phys)
and $a_{aa}=2$ (\krk)
which dominates the visible differences between the schemes.%
\footnote{Note that for \dis $a_{gg}=0$, which is reflected in its similarity to \msbar, \mpos($\delta$) in \cref{fig:NspacePgg}.}
The starkly differing behaviour of the \dis scheme in the $gq$ case
is due to the defining constraint on the gluon,
which generates unphysical terms arising from $p_{qq}$.
This includes $\mathcal{D}_1$ terms (i.e., $\log^2 N$),
within $\rK_{gq}^{\msbar\to\dis}$ and thus $\rP_{gq}^{\dis}$, as in \cref{eq:PDISgq}.
This also applies in the non-singlet sector in \cref{fig:NspacePq}, where for
$P_{q_iq_k}$, $P_{q_i\overline{q}_i}$ and $P_{q_i\overline{q}_k}$
the miscancellation between $qg$ and $gq$ transformations in the
\dis scheme introduces terms proportional to $\log^2(1-x)$
($N^{-1} \log^2 N$ terms in Mellin space), which decay
slower as $N\to \infty$ than the leading term in the general scheme
$\propto (1-x) \log(1-x)$,
\begin{align}
\cf \tr (a_{gq}-a_{qg}) \frac{\log N}{N^2},
\end{align}
and significantly slower than the leading term in the \krk or \phys schemes, for which $a_{qg}=a_{gq}$.

\section{Asymptotic analysis and resummation implications}
\label{sec:asymptotics}

The possible magnitude of perturbative corrections to the DGLAP splitting functions,
and therefore scheme-dependence, is $x$-dependent.
Towards the interior of the unit interval $(0,1)$,
the perturbative expansion of the splitting functions is relatively well-behaved, in any scheme:
mixing between the channels introduces modifications that are bounded multiples of
the splitting functions at lower orders, and the
direct modification proportional to the beta-function (see \cref{eq:DGLAPFStrans})
is regular in the interior of the interval 
under the assumptions of
\cref{eq:tableDecomp}.
The NLO corrections may be large, but are bounded,
and pointwise perturbative.

In the limits approaching 0 and 1, this is no longer the case,
due to the introduction of divergent contributions proportional to
$(\log^k x)/x$ and $(\log^k (1-x)) / (1-x)$ respectively.
Factorisation scheme transformations can change both the coefficient and the degree $k$ of the leading
non-zero term in each limit, yielding unbounded differences between schemes.

Since the behaviour of the splitting functions in these limits can be singular,
the perturbative suppression of higher orders in $\alphas$ is
overwhelmed by the growth of the singular terms.
This motivates the reorganisation of the series into distinct, appropriate, expansion
parameters, and its eventual resummation.

As $x \to 1$, the splitting functions probe
soft-gluon singularities, giving divergent contributions
$\mathcal{D}_k$ (equivalently, $\log^k N$ in Mellin space).
The singlet matrix diagonalises in the usual flavour-diagonal channels,
and, in the \msbar scheme, the coefficient of the leading term $\mathcal{D}_0$ at each order
is given by the cusp anomalous dimension, $\Gamma_{\mathrm{cusp}}$, arising from the UV structure of cusped Wilson loops.
These obey Casimir scaling (up to four loops \cite{Grozin:2017css,Henn:2019swt,vonManteuffel:2020vjv}, 
where it is violated due to the emergence of non-quadratic Casimirs).
In a general scheme, this need not remain, and is conditional on the form of the transformation kernels.%
\footnote{As a result, preserving Casimir scaling induces a constraint on the \gen scheme coefficients.}

As $x \to 0$, the dominant singularity is given by the $1/x$ pole of $\rP_{gg}$,
governing the growth of the gluon distribution at small-$x$.
Additional powers of $\log^k x$ 
are typically resummed using the BFKL equation \cite{Kuraev:1976ge,Balitsky:1978ic}
within the high-energy or $k_T$-factorisation framework~\cite{Collins:1991ty,Catani:1990eg,Catani:1990xk,Kotko:2015ura,Hentschinski:2017ayz,vanHameren:2022mtk},
which shifts the pole at $N=1$ in Mellin space to $N=1 + \omega_P$, where $\omega_P$ is the
pomeron intercept (approximately 0.5),
leading to a gluon distribution $f_g \sim x^{-(1+\omega_P)}$ at small-$x$.

In a general scheme, the degree of singularity in both limits can deviate from the familiar \msbar behaviour.
As we shall see, the dominant singular corrections arising in alternative schemes are proportional to $\beta^{(2)}$,
and admit a natural interpretation as artefacts of expanding the coupling $\alphas(\mu f(x))$ of an
$x$-dependent renormalisation scale $\mu_{\mathrm{eff}} = \mu f(x)$ as a series in $\alphas(\mu)$.

\subsection{\texorpdfstring{Large-$x$ ($N \to \infty$)}{Large-x (N → ∞)}}
\label{sec:disc_zto1}
The asymptotics of the \msbar splitting functions in the $x\to 1$
region are known to all orders
\cite{Korchemsky:1988si,Korchemsky:1991zp,Albino:2000cp,Dokshitzer:2005bf,Vogt:2010cv}.
The dominant and subdominant contributions
to the diagonal splitting functions in the \msbar
scheme
are given, to all orders $k$, by
\begin{align}
    \label{eq:Paalargex}
    \rP^{\msbar(k)}_{aa}(x) & {} \sim A^{(k)}_a \left[\frac{1}{1-x}\right]_+ + B_{a}^{(k)} \, \delta(1-x)
    + \order{\log(1-x)}
\end{align}
for $a\in\{q,g\}$, which corresponds in Mellin space to 
\begin{align}
    \notag
    \mathcal{M}\left[\rP^{\msbar(k)}_{aa}\right](N) & {} \sim - A^{(k)}_a \log N 
    + ( B_{a}^{(k)} - \egamma A_{a}^{(k)} )
    + \order{\frac{1}{N}\log N}
\end{align}
where $A_q, A_g$ are the quark and gluon
cusp anomalous dimensions,
related by Casimir scaling (at the orders of interest),
\begin{align}
A_g = \frac{\ca}{\cf} A_q.
\end{align}
Correspondingly, as $x\to 1$, the \msbar off-diagonal splitting functions have 
asymptotic behaviour
$\order{\log^{2(k-1)}(1-x)}$, or equivalently
$\order{N^{-1} \log^{2(k-1)}N}$ in Mellin space.

Considering the transformations in \cref{eq:DGLAPFStrans},
in the off-diagonal case, for all schemes the commutator in \cref{eq:DGLAPFStrans_comm} gives, for example
\begin{align}
\bigl[\mathbb{K}^{(1)}, \mathbb{P}^{(1)}\bigr]_{qg} = {} &
    \left(\mathrm{K}^{(1)}_{qq} - \mathrm{K}^{(1)}_{gg}\right)\otimes \mathrm{P}^{(1)}_{qg}
    + \mathrm{K}^{(1)}_{qg} \otimes (\mathrm{P}^{(1)}_{gg} - \mathrm{P}^{(1)}_{qq}).
\end{align}
In Mellin space, both terms are suppressed by factors of $1/N$ at large $N$ from $\mathrm{P}^{(1)}_{qg}$ and $\mathrm{K}^{(1)}_{qg}$ respectively (as $p_{qg}$ is regular as $x\to 1$).
The remaining part of the transformation in \cref{eq:DGLAPFStrans},
$(\beta^{(2)}/(2\pi)) \rK_{qg}^{\msbar\to\rFS}$, is suppressed because of the form assumed in \cref{eq:genschemeKqg}.
The large-$N$ behaviour is therefore unchanged from the \msbar case, as can be seen in \cref{fig:xspacePqg,fig:NspacePqg}.%
\footnote{Due to the departure of the \dis scheme kernels from the form given in \cref{eq:genschemeKgq}
(by containing $\mathcal{D}_{0,1}$ terms),
this conclusion is false for the \dis scheme, as noted in \cite{White:2005wm}.}
For $\rP_{gq}$, the scheme-dependent terms contribute at large-$x$ as $\log^2(1-x)$
and so diverge as $x\to 1$ but vanish as $N\to \infty$
(and hence the schemes generally appear similar at large-$N$ in \cref{fig:NspacePgq} despite diverging in \cref{fig:xspacePgq}).

The rate of decay as $N \to \infty$ is governed by the first neglected term
(i.e.\, the next term in the expansion),
proportional to $\log^2(1-x)$, whose coefficients can be read off from
\cref{eq:PGENgq,eq:PGENqig}.

For diagonal splittings, the commutator in \cref{eq:DGLAPFStrans_comm}
simplifies:
\begin{align}
    \bigl[\mathbb{K}^{(1)}, \mathbb{P}^{(1)}\bigr]_{qq}
& {} =
    \mathrm{K}^{(1)}_{qg} \otimes \mathrm{P}^{(1)}_{gq}
    -
    \mathrm{P}^{(1)}_{qg} \otimes \mathrm{K}^{(1)}_{gq}
= - \bigl[\mathbb{K}^{(1)}, \mathbb{P}^{(1)}\bigr]_{gg},
\end{align}
so we can consider the two diagonal channels together.
If we consider this commutator in Mellin space
the off-diagonal splitting-functions are asymptotically vanishing,
as are the off-diagonal kernels in \cref{eq:genschemeKqg,eq:genschemeKgq}.%
\footnote{Of the schemes considered in
\cite{Delorme:2025teo}, this is violated only by the \dis scheme kernel $\mathrm{K}^{\msbar\to\dis(1)}_{gq}$.}
In this case, the commutator therefore does not modify the asymptotic $x\to 1$ ($N\to\infty$) behaviour of the \msbar splitting functions.
Any change is therefore attributable to the remaining part of the transformation in \cref{eq:DGLAPFStrans},
\begin{align}
    (\beta^{(2)}/(2\pi)) \rK_{aa}^{\msbar\to\rFS}.
\end{align}
The $x\to 1$ divergent terms in \cref{eq:genschemeKqq,eq:genschemeKgg}
are
\begin{align}
\rK_{aa}^{\msbar\to\rFS}
& {} \sim
C_a
\left[ 2 a_{aa} \mathcal{D}_1 - b_{aa} \mathcal{D}_0 - \Delta_{aa} \, \delta(1-x) \right]
+ \order{\log(1-x)},
\end{align}
again for $a\in\{q,g\}$ and with $C_q = \cf$, $C_g = \ca$.
Therefore, the leading $x\to 1$ behaviour of the 
transformed splitting functions is given by
\begin{align}
    \label{eq:PaaFS_largez}
    \rP^{\rFS(2)}_{aa}(x) & {} \sim
    C_a \frac{\beta^{(2)}}{\pi} a_{aa} \mathcal{D}_1
    +
    \left(A^{(2)}_a - C_a \frac{\beta^{(2)}}{2\pi} b_{aa} \right) \mathcal{D}_0
    + \left( B_{a}^{(2)} - C_a \frac{\beta^{(2)}}{2\pi} \Delta_{aa} \right) \delta(1-x)
    + \order{\log(1-x)}.
\end{align}
This behaviour can be verified in the scheme-specific calculations above.
In Mellin space, this corresponds to
\begin{align}
\label{eq:MPFSaaN}
    \mathcal{M}\left[\rP^{\rFS(2)}_{aa}\right](N) & {} \sim
     C_a \frac{\beta^{(2)}}{2\pi} a_{aa} \log^2 N
    +
    \left(C_a \frac{\beta^{(2)}}{2\pi} (b_{aa}+2a_{aa}\egamma) - A^{(2)}_a \right) \log N
    + \biggl( B_{a}^{(2)} - C_a \frac{\beta^{(2)}}{2\pi} \Delta_{aa} \biggr)
    \\ \notag & {}
    +\egamma\left(C_a \frac{\beta^{(2)}}{2\pi} b_{aa} - A^{(2)}_a \right)
    + C_a \frac{\beta^{(2)}}{2\pi} a_{aa}\left(\egamma^{2}+\frac{\pi^2}{6}\right)
    \biggr)
    + \order{\frac{1}{N}\log N}.
\end{align}
Note that, for transformations of this form, the Casimir scaling relationship between
the quark and gluon diagonal splitting-functions is preserved, provided that
$a_{qq} = a_{gg}$ and $b_{qq} = b_{gg}$.%
\footnote{For the schemes considered in \cite{Delorme:2025teo}, this is the case for the \krk, \phys, and \pos-family schemes (for which $a_{qq} = a_{gg} = 2, 1, 0$ respectively; $b_{qq}=b_{gg}=0$).
Imposing the preservation of this Casimir scaling relationship can in principle be used to reduce the dimensionality of the parameter space of factorisation schemes.}

The leading double-logarithmic terms ($\propto C_a a_{aa}$) in \cref{eq:MPFSaaN} correspond to threshold logarithms
that are moved by the factorisation-scheme transformation
from the coefficient functions into the PDFs, where DGLAP evolution will now resum them,
as discussed further in \cref{sec:asymptotics_resummation}.
They can be interpreted as implying an effective evolution scale, discussed in \cref{sec:asymptotics_effscales}.

\subsection{\texorpdfstring{Small-$x$ ($N \to 1$)}{Small-x (N → 1)}}
\label{sec:disc_zto0}

Near $N=1$, the asymptotic behaviour is driven by the coefficient and degree of singularity of the leading terms as $x \to 0$.
The leading small-$x$ behaviour at leading order is
\begin{align}
    \rP_{qq}^{\rFS(1)} & {} \sim 0
    & \rP_{qg}^{\rFS(1)} & {} \sim 0 \\
    \rP_{gq}^{\rFS(1)} & {} \sim 2 \cf \frac{1}{x}
    & \rP_{gg}^{\rFS(1)} & {} \sim 2 \ca \frac{1}{x},
\end{align}
which corresponds to a leading singularity $\sim 1/(N-1)$ in Mellin space,
since
\begin{align}
\label{eq:Mzm1logkz}
    \mathcal{M}[ z^{-1} \log^k z ](N) &= (-1)^{k} \frac{k!}{(N-1)^{k+1}}.
\end{align}
Contributions in $x$ space more singular than this would shift
the location of the rightmost pole in Mellin space,
rather than merely changing its degree (through additional logarithmic powers
in the numerator) or residue.

At NLO, in the \msbar scheme, the leading behaviour is
\begin{align}
    \rP_{qq}^{\msbar(2)} & {} \sim \cf \nf \tr \frac{40}{9x} &
    \rP_{qg}^{\msbar(2)} & {} \sim \ca \nf \tr \frac{40}{9x} \\
    \rP_{gq}^{\msbar(2)} & {} \sim \cf \left( \ca - \frac{40}{9} \nf \tr \right) \frac{1}{x} &
    \rP_{gg}^{\msbar(2)} & {} \sim \nf \tr \left( \frac{4}{3}\cf - \frac{46}{9} \ca \right) \frac{1}{x},
\end{align}
with the same leading singular term.%
\footnote{The accidental cancellation of the leading-logarithmic terms 
for spacelike kernels at low perturbative orders in the \msbar scheme has long been known, see for example \cite{Catani:1994sq,Ball:1995vc,Bonvini:2018xvt}.}

This leading singular behaviour admits modifications in a general scheme,
arising from convolutions with as-singular or more-singular terms,
of the form given in \cref{eq:Mzm1logkz},
or directly from the contribution proportional
to $\beta^{(2)} \rK_{ab}^{\msbar \to \rFS}$.
In the latter case,
since $p_{gq}$ and $p_{gg}$ contain leading $1/x$ terms, they
lead to direct modifications of the splitting functions by
$ \propto c_{gq} \log(x)/x$.
This is already more singular than the leading \msbar behaviour,
for $c_{gq}$ or $c_{gg}$ non-zero.
In the former case, we may again consider the $x$-space convolutions
as products in Mellin space, to identify the leading terms arising from 
the commutator.
Where the same $c_{gq}$ and $c_{gg}$ $\log(x)/x$ terms 
(two inverse powers of $(N-1)$)
are convolved with
leading-order splitting functions $\propto 1/x$, 
(an additional inverse power)
in $x$ space,
arising from $\rP_{gq}^{(1)}$ and $\rP_{gg}^{(1)}$
they become double-logarithmic
and the commutator structure generates a difference of coefficients,
\begin{align}
    \rP_{gq}^{\gen(2)} \ni 2 \ca \cf (c_{gg}-c_{gq}) \frac{\log^2 x}{x}.
\end{align}
This is the most singular term that can arise from a scheme transformation,
and is essentially
an artefact of a miscancellation between inconsistent treatments of high-energy gluon
radiation between the $gg$
and $gq$ scheme transformations.%
\footnote{As shown in \cref{tab:FScoeffs}, this is the case for the \mpos family of schemes, and can be seen in their asymptotic behaviour
in \cref{fig:xspacePgq,fig:NspacePgq}, and the \aversa scheme,
which in the context of this leading term may be considered to have $c_{gg}=0,c_{gq}=1$.}

Since the leading-order $\rP_{qb}^{(1)}$ are finite as $x \to 0$,
the leading term in $\rP^{\gen(2)}_{qq}$ arises from the $- \gamma_{qg}^{(1)}(1) \rK_{gq}$
in the commutator,
and in $\rP^{\gen(2)}_{qg}$ likewise from the $- \gamma^{(1)}_{qg} (1) \rK_{gg}$ term,
since
\begin{align}
    \gamma_{qg}^{(1)}(1) = 2 \nf \tr \int_0^1 \pqg \; \dd z = \frac{4}{3} \nf \tr .
\end{align}
These combine with the leading $2/x$ singularity in $p_{gq}$ and $p_{gg}$ respectively to give
\begin{align}
    \rP_{qq}^{\gen(2)} & {} \ni \frac{8}{3} \cf \nf \tr c_{gq} \frac{\log x}{x}, \\
    \rP_{qg}^{\gen(2)} & {} \ni \frac{8}{3} \ca \nf \tr c_{gg} \frac{\log x}{x}.
\end{align}

Finally, in the $gg$ case, as in \cref{sec:disc_zto1}, the diagonal contributions from the commutator cancel,
leaving a leading term arising from $\rK_{gq} \gamma_{qg}^{(1)}(1)$, originating from the same source as those
for $qq$ and $qg$ above. This combines with the beta-function term coming directly from the
transformation kernel $\rK_{gg}^{\msbar \to \rFS}$, discussed above, to give
\begin{align}
    \rP_{gg}^{\gen(2)} & {} \ni -\left[\ca \frac{\beta^{(2)}}{\pi} c_{gg} + \frac{8}{3} \cf \nf \tr c_{gq} \right]
    \frac{\log x}{x} .
\end{align}

Treating the subleading terms carefully, the full leading small-$x$ behaviour for the generalised scheme is
\begin{align}
    \rP_{qq}^{\text{GEN}(2)} \sim {} & \cf \nf \tr
    \biggl[ \frac{8}{3}c_{gq} \frac{\log x}{x} 
    + \frac{2}{9x} \bigl( 20 + 13(-a_{qg} + c_{gq} + c_{qg}) \bigr) \biggr]
     + \order{\log^2 x} ,
     \\
    \rP_{qg}^{\text{GEN}(2)} \sim {} & \ca \nf \tr \biggl[ \frac{8}{3} c_{gg} \frac{\log x}{x} 
    + \frac{2}{9x} \bigl( 20 + 13(- a_{qg} + c_{gg} + c_{qg}) \bigr) \biggr]
     + \order{\log^2 x} ,
     \\
    \rP_{gq}^{\text{GEN}(2)} \sim {} & 2 \ca \cf (c_{gg} - c_{gq}) \frac{\log^2 x}{x}
    + \cf \left[ 3 \ca c_{gg} - \frac{8}{3} \nf \tr c_{gq} \right] \frac{\log x}{x}
    \\ \notag
    & {} \qquad
    + \frac{\cf}{x}
    \biggl[
    \ca
    \left(\frac{2\pi^2}{3}(-a_{gg}+a_{gq}+c_{gg}+c_{gq}) + 1 + \frac{1}{18}(134 a_{gg} - 45 a_{gq} - 71 c_{gg} - 71 c_{gq} - 36 \Delta_{gg})\right)
    \\ \notag
    & {} \qquad
    + \cf \left(-\frac{2\pi^2}{3}(c_{gq}+c_{qg})+\frac{1}{2}(-7a_{qq}+5c_{gq}+5c_{qq}) + 2 \Delta_{qq}\right)
    - \frac{40}{9} \nf \tr
    \biggr]
     + \order{\log^2 x} ,
    \\
    \rP_{gg}^{\text{GEN}(2)} \sim {} &
    -\left[\ca \frac{\beta^{(2)}}{\pi} c_{gg} + \frac{8}{3} \cf \nf \tr c_{gq} \right]
    \frac{\log x}{x} 
    + 
    \frac{2}{9x} \nf \tr \biggl[ 
    \cf \bigl( 6 + 13 (a_{qg} - c_{gq} - c_{qg}) \bigr)
    - 23 \ca
    \biggr]
     + \order{\log^2 x} .
\end{align}

As shown in \cref{tab:FScoeffs}, for the named schemes considered in \cref{sec:calcres}, 
\begin{align}
    c_{ab} = \begin{cases}
        1 & \krk \\
        0 & \phys
    \end{cases}
\end{align}
which explains the absence of $\log(x)/x$ terms from
\cref{eq:PPHYSqiqi,eq:PPHYSqiqk,eq:PPHYSqiqbi,eq:PPHYSqig}
and why the \phys scheme resembles the \msbar scheme in the $N\to 1$
limit in \cref{fig:NspaceP,fig:NspacePq}.
The \krk scheme has different asymptotic behaviour
resulting in its absorption of universal small-$x$ terms from
coefficient functions into the PDFs.

Again identifying the eigenvalues of the singlet coupling matrix $\bm{\mathrm{\gamma}}$,
this time expanding in the small-$x$ limit,
gives eigenvalues $\lambda_\pm$ ($\lambda_- < \lambda_+$),
with perturbative expansion
\begin{align}
   \lambda_{\pm} = \left(\frac{\alphas}{2\pi}\right) \lambda_{\pm}^{(1)} + \left(\frac{\alphas}{2\pi}\right)^2 \lambda_{\pm}^{(2)} ,
\end{align}
given in \cref{eq:lambdap1_smallx,eq:lambdam1_smallx,eq:lambdap2_smallx,eq:lambdam2_smallx},
\begin{align}
    \label{eq:lambdap1_smallx}
    \lambda_+^{(1)}(N) & {} =
    \frac{2\ca}{N-1} + \biggl(- \frac{11}{3}\ca - \frac{\beta^{(2)}}{2\pi} + \frac{4 \cf \nf \tr}{3 \ca} \biggr) + \order{(N-1)^1} ,
    \\ \label{eq:lambdap2_smallx}
    \lambda_+^{(2)}(N) & {} =
    \frac{2\ca c_{gg}}{(N-1)^2} \frac{\beta^{(2)}}{2\pi}
    \\ \notag & {}
    + \frac{1}{N-1} \frac{2 \nf \tr}{9 \ca^2} \biggl( - 23 \ca^3 + 26 \ca^2 \cf + \left( 24 \ca^2 \cf - 4 \ca \cf \nf \tr \right) (c_{gg}-c_{gq}) - 16 \cf^2 \nf \tr c_{gq} \biggr)
    + \order{1} ,
    \\ \label{eq:lambdam1_smallx}
    \lambda_-^{(1)}(N) & {} =
    -\frac{4 \cf \nf \tr}{3 \ca} + \order{(N-1)^1} ,
    \\ \label{eq:lambdam2_smallx}
    \lambda_-^{(2)}(N) & {} =
    \frac{1}{N-1} \frac{8 \cf \nf \tr}{9 \ca^2}\biggl( (-6\ca^2 + \ca \nf \tr)(c_{gg}-c_{gq}) + 4 \cf \nf \tr c_{gq} \biggr) + \order{1} .
\end{align}
As can be seen, the degree of singularity of the larger eigenvalue is scheme-dependent,
but the $gq$-terms $\sim \log^2(x)/x$ do not contribute to this, since they combine with suppressed
$qg$-terms in the determinant.
The smaller eigenvalue $\lambda_-$, which is finite in the \msbar scheme, is singular in general, but less so
than the larger eigenvalue unless $c_{gg}=0$ and $c_{gq} \neq 0$.

\subsection{Interpretation: effective evolution scales}
\label{sec:asymptotics_effscales}
The dominant contributions to the splitting functions identified in
\cref{sec:disc_zto0,sec:disc_zto1} share a common feature:
in both cases, they arise from the direct term proportional to the beta-function
in \cref{eq:DGLAPFStrans}, rather than the commutator with the leading-order splitting functions,
and are proportional to the leading-order splitting function, multiplied by
a logarithm of a function of $x$.
Such terms have a natural physical interpretation.

The solution to the beta-function evolution of $\alphas(\mu)$, 
\cref{eq:betafunction}, truncated to NLO, is
\begin{align}
    \alphas(\mu) = \alphas (\mu_0) \left[
    1 + \frac{\alphas(\mu_0)}{2\pi} \frac{\beta^{(2)}}{\pi} \log\frac{\mu}{\mu_0}
    + \order{\alphas^2}
    \right].
\end{align}
Evolution according to the leading-order splitting functions, if associated with a different
effective evolution scale $\mu_{\mathrm{eff}}$,
therefore generates $\order{\alphas}$ terms arising from the
shift in the coupling,
\begin{align}
\label{eq:alphasevol}
    \alphas(\mu_{\mathrm{eff}})\,
    \rP^{(1)}(x)
    =
    \alphas(\mu) \,
    \rP^{(1)}(x)
    \left[ 1
    +
    \frac{\alphas(\mu)}{2\pi} \frac{\beta^{(2)}}{\pi} \log\frac{\mu_{\mathrm{eff}}}{\mu}
    \right].
\end{align}
Imposing an expansion in $\alphas(\mu)$, rather than the $x$-dependent scale $\mu_{\mathrm{eff}}$, induces additional $x$-dependent terms proportional to the beta-function and the leading-order splitting functions.
These are precisely the leading terms identified in \cref{sec:disc_zto0,sec:disc_zto1}.

Recognising the leading large-$x$ behaviour of the diagonal splitting functions \cref{eq:Paalargex} as, at LO,
\begin{align}
    \rP^{(1)}_{aa}(x) \sim \frac{2 C_{a}}{1-x},
\end{align}
we can interpret the leading term in \cref{eq:PaaFS_largez} as (separating the numerator and denominator of $\mathcal{D}_1$),
\begin{align}
    C_a \frac{\beta^{(2)}}{\pi} a_{aa} \mathcal{D}_1
    = \left[ \left(\frac{2 C_a}{1-x}\right) \left(\frac{\beta^{(2)}}{\pi} \log(1-x)^{\frac{1}{2}a_{aa}} \right) \right]_+,
\end{align}
that is, corresponding to the terms that would be generated by an effective evolution scale
\begin{align}
    \mu_{\mathrm{eff}}
    = \mu (1-x)^{\frac{1}{2} a_{aa}}.
\end{align}
For the values of $a_{aa}$ adopted in the schemes of \cref{tab:FScoeffs},
$a_{aa} \in \{ 0,1,2 \}$,
this corresponds to evolving the large-$x$ splitting functions at
the effective scale
\begin{align}
    \mu_{\mathrm{eff}}
    =
    \begin{cases}
        \mu & \msbar, \mpos(\delta) \\
        \mu \sqrt{1-x} & \aversa, \dis \text{ ($qq$ only)}, \phys \\
        \mu (1-x) & \krk.
    \end{cases}
\end{align}
This is reminiscent of the scale hierarchy of soft-collinear effective theory (SCET), matching the
hard,
collinear/jet-like,
and soft scales respectively \cite{Becher:2006nr,Becher:2006mr,Becher:2007ty}.

As in the large-$x$ case, the leading small-$x$ behaviour can be interpreted as corresponding to an effective evolution scale.
Analogously to \cref{eq:alphasevol}, the larger, gluon-dominated, eigenvalue $\lambda_+$, singular at LO as
\begin{align}
    \lambda_+^{(1)}(x) \sim \frac{2\ca}{x}
\end{align}
is the factor appearing alongside the beta-function coefficient and scale-logarithm in the leading behaviour of the NLO correction,
\begin{align}
    \lambda_+^{(2)}(x) \sim
    \left( \frac{2\ca}{x} \right) \left( \frac{\beta^{(2)}}{\pi}
    \log x^{-\frac{1}{2}c_{gg}}\right),
\end{align}
implying an effective high-energy evolution scale of
\begin{align}
    \mu_{\mathrm{eff}}
    = \mu x^{-\frac{1}{2}c_{gg}}.
\end{align}
For the values of $c_{gg}$ adopted in the schemes of \cref{tab:FScoeffs},
$c_{gg} \in \{ 0,1 \}$,
this corresponds to evolving the small-$x$ gluon at
\begin{align}
    \mu_{\mathrm{eff}} = \frac{\mu}{\sqrt{x}}
\end{align}
for the \krk scheme, for which $c_{gg} = 1$.

The synthesis of these two limits is that the large-logarithmic corrections, taken together, correspond to an effective scale
of the form
\begin{align}
    \label{eq:mueff_interp}
    \mu_{\mathrm{eff}}^2 = \mu^2 \frac{(1-x)^{a_{aa}}}{x^{c_{aa}}}.
\end{align}
This interpolates between the two limits,
and gives a physical interpretation to the family of $\gen$ scheme parameters.
Special cases of scales of this form include, for example, the hierarchy of scales that arises from
the angular-ordering of coherent evolution \cite{Ciafaloni:1987ur,Catani:1989yc,Catani:1989sg,Kwiecinski:1996td}
\begin{align}
  k_{\rT}^2 < Q^2 \frac{1-x}{x},
\end{align}
the invariant mass of the final hadronic state in DY and DIS (the SCET `jet scale' $\mu_J$ \cite{Becher:2006mr}),
\begin{align}
    M_X^2 
    \approx Q^2 \frac{1-x}{x} \underset{x\to1}{\sim} Q^2 (1-x) \sim \mu_J^2,
\end{align}
or the threshold-resummation \cite{Sterman:1986aj,Catani:1989ne,Forte:2002ni} (and SCET \cite{Becher:2006mr}) `soft scale',
\begin{align}
  k_{\rT}^2 \lesssim \mu_s^2 \sim Q^2 (1-x)^2.
\end{align}

Indeed, revisiting \cref{eq:DGLAPFStrans}, the beta-function term can be identified
as the perturbative correction that emerges from a scale choice
\begin{align}
    \mu_{\mathrm{eff}}
    =
    \kappa_{ab}(x) \mu,
\end{align}
where
\begin{align}
    \label{eq:effscaleexp}
    \kappa_{ab}(x) = \exp\left[ \frac{1}{2 \, C_{ab} \, p_{ab}(x)} \; \rK_{ab}^{\msbar\to \rFS} (x)\right]
\end{align}
where here $C_{ab}$ is the colour factor of the leading-order splitting function
$P_{ab}^{(1)}$.
(This identification is perhaps unsurprising, considering the term's origins as the
logarithmic derivative of $\alphas$.)
In the small- and large-$x$ limits, the dominant terms of the scheme-transformation kernels,
controlled by the $c$ and $a$ parameters respectively in the \gen scheme,
are logarithmic, and we recover \cref{eq:mueff_interp}.
This implies that the dominant effect of evolution in alternative schemes can be obtained
by modifying the scale of $\alphas$ at LO in existing evolution codes,%
\footnote{Note that for Mellin-space evolution codes, promoting the scale of the running coupling to a function of $x$
implies that $\alphas$ is no longer constant in Mellin space and would need to be included in the transformation integral.}
which will also resum higher-order terms associated with such an effective scale.

\subsection{Implications for resummation}
\label{sec:asymptotics_resummation}

Considering for example any non-singlet distribution,
the corresponding DGLAP equations decouple,
i.e.\ in Mellin space,
\begin{align}
	\label{eq:dglap_ns}
	\frac{\partial}{\partial (\log \mu^2)}
    \mathcal{M}[
    f_{\mathrm{NS}}^{\rFS}(\mu)](N)
	  = {} &
    \gamma_{\mathrm{NS}}(N) \;
	\mathcal{M}[f_{\mathrm{NS}}^{\rFS} (\mu)](N).
\end{align}
This has a closed-form solution, which we write
in terms of $\tilde{N} = N e^{\egamma}$ and expanded in $N$:
\begin{align}
    \notag
    & \mathcal{M}[
    f_{\mathrm{NS}}^{\rFS}(\mu)](N)
	{} =
     \mathcal{M}[
    f_{\mathrm{NS}}^{\rFS}(\mu_0)](N)
    \left( \frac{\alphas(\mu)}{\alphas(\mu_0)} \right)^{\frac{2\pi}{\beta^{(2)}}\gamma^{(1)}(N)}
    \exp\left[
    \left(\frac{\gamma^{\rFS(2)}(N)}{\beta^{(2)}} - \frac{\gamma^{(1)}(N)\beta^{(3)}}{(\beta^{(2)})^2} \right)
    \left( \alphas(\mu) - \alphas(\mu_0) \right)
    + \order{\alphas^2}
    \right]
    \\ 
    & {} =
     \mathcal{M}[
    f_{\mathrm{NS}}^{\rFS}(\mu_0)](N)
    \left( \frac{\alphas(\mu)}{\alphas(\mu_0)} \right)^{\frac{2\pi}{\beta^{(2)}}\gamma^{(1)}(N)}
    \exp\Biggl[
    \left( \frac{\alphas(\mu) - \alphas(\mu_0)}{2\pi} \right)
    \Biggl\{
    C_a a_{aa} \log^2 \tilde{N}
    \\ \notag  & {} \qquad
    + \left(
    C_a b_{aa} 
    - \frac{2\pi }{\beta^{(2)}}
    \left(A_a^{(2)}  -  A_a^{(1)} \frac{\beta^{(3)}}{\beta^{(2)}} \right)
    \right)
    \log \tilde{N}
    + C_a a_{aa} \frac{\pi^2}{6}
    + \frac{2\pi}{\beta^{(2)}} 
    \left( B_a^{(2)}
    - B_a^{(1)} \frac{\beta^{(3)}}{\beta^{(2)}} \right)
      - C_a \Delta_{aa}
    + \mathcal{O}\!\left(\frac{\log N}{N}\right)\Biggr\}
    \\
	\label{eq:dglapsol_ns}
    & {} =
    \mathcal{M}[
    f_{\mathrm{NS}}^{\rFS}(\mu_0)](N)
    \left( \frac{\alphas(\mu)}{\alphas(\mu_0)} \right)^{\frac{2\pi}{\beta^{(2)}}\gamma^{(1)}(N)}
    \exp\Biggl[
    \left( \frac{\alphas(\mu) - \alphas(\mu_0)}{2\pi} \right)
    \Biggl\{
    C_a a_{aa} \log^2 N
    \\ \notag  & {} \qquad
    + \left(
    2 C_a a_{aa}\egamma + C_a b_{aa} 
    - \frac{2\pi }{\beta^{(2)}}
    \left(A_a^{(2)}  -  A_a^{(1)} \frac{\beta^{(3)}}{\beta^{(2)}} \right)
    \right)
    \log N
    \\ \notag  & {} \qquad
    + C_a a_{aa}\Big(\egamma^2+\frac{\pi^2}{6}\Big)
    + \left(C_a b_{aa} - \frac{2\pi}{\beta^{(2)}}
    \left( A_a^{(2)} - A_a^{(1)} \frac{\beta^{(3)}}{\beta^{(2)}} \right)
    \right) \egamma
    + \frac{2\pi}{\beta^{(2)}} 
    \left( B_a^{(2)}
    - B_a^{(1)} \frac{\beta^{(3)}}{\beta^{(2)}} \right)
    - C_a \Delta_{aa}
    \\ \notag & {} \qquad
    + \order{\frac{1}{N}\log N}
    \Biggr\}
    + \order{\alphas^2}
    \Biggr]
\end{align}

\begin{figure*}[tpb]
    \centering
    \includegraphics[width=\linewidth]{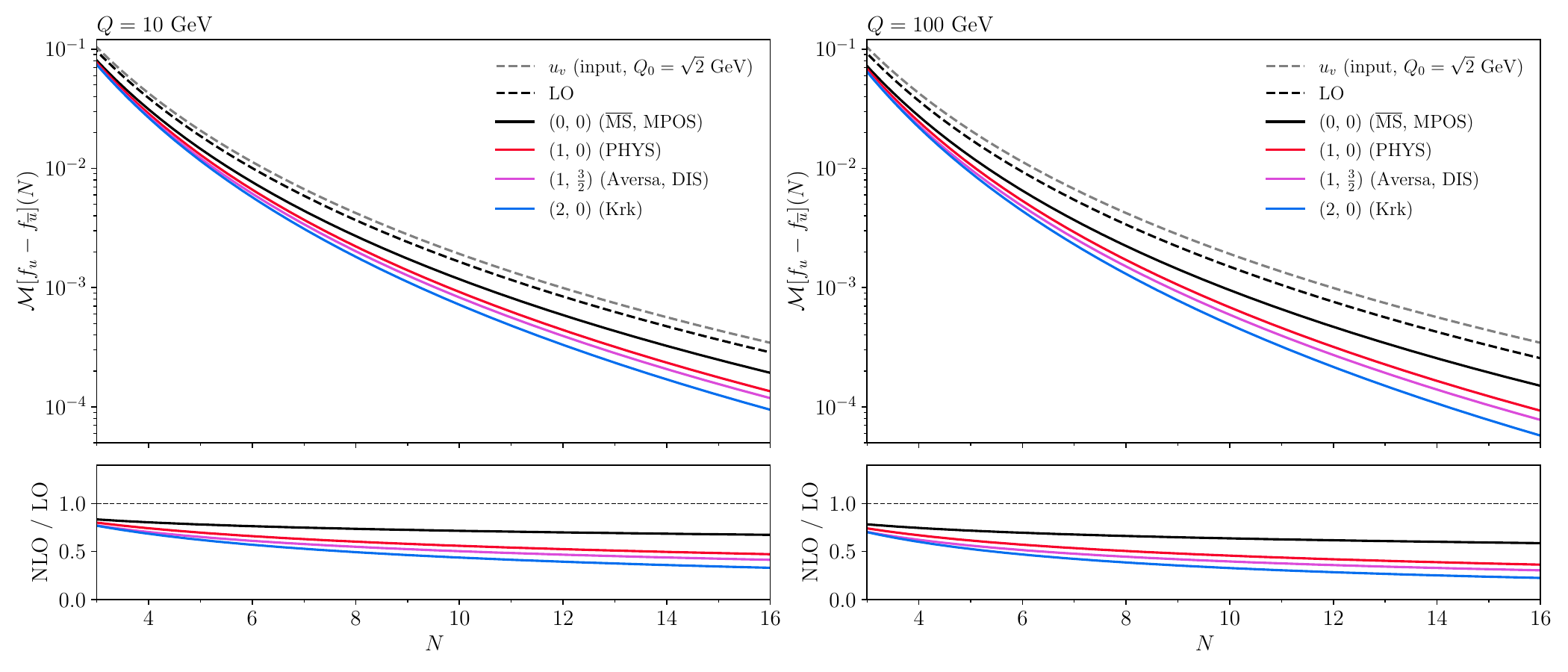}
    \caption{Non-singlet evolution of the prototypical up-quark valence distribution
    from the Les Houches PDF benchmarking exercise \cite{Giele:2002hx}
    using
    the truncated large-$N$ approximation given in \cref{eq:dglapsol_ns},
    from input scale $Q_0 = \sqrt{2}$ GeV to $Q=10,100$ GeV.
    The values of $a_{qq}$ and $b_{qq}$ used are shown, alongside the named schemes that use those values (see \cref{tab:FScoeffs}).
    Here, we fix the value of $\Delta_{qq}$ corresponding to momentum conservation
    for the included components according to \cref{eq:GENDeltaqq}, i.e.\
    $\Delta_{qq} = \frac{121}{36} a_{qq} + b_{qq}$,
    which differs from the value for the named schemes (due to their inclusion of non-zero $P(z)$).
    The \gen-scheme large-$N$ approximation of the named-scheme evolution kernels is reasonable
    for the $N$-values shown.
    \label{fig:GENvalenceevol}}
\end{figure*}

As a consequence, in the general case of $a_{aa}$ non-zero, the PDF evolution at large-$x$ is expected to be substantially altered from the \msbar case.
This is due to the presence of double-logarithmic terms absent
in the \msbar scheme, leading to suppression at large $N$ for positive $a_{aa}$.

To quantify this, in \cref{fig:GENvalenceevol}, we plot the solution of \cref{eq:dglapsol_ns} for different realisations of the \gen scheme. As expected, the largest deviation from \msbar is obtained for the \krk-like variant of the \gen scheme,
with large-$N$ suppression relative to \msbar comparable to the size of the \msbar NLO suppression relative to LO.%
\footnote{Studying the magnitude of this suppression for $x$-space distributions would require $x$-space evolution or Mellin inversion, which we defer to future work.}

For the coupled singlet--gluon system, which evolves as
\begin{align}
    \frac{\partial}{\partial(\log \mu^2)} 
    \begin{pmatrix}
    \mathcal{M}[\Sigma^{\rFS}(\mu)](N) \\
    \mathcal{M}[f_{g}^{\rFS}(\mu)](N)
    \end{pmatrix}
	{} & =
    \bm{\mathrm{\gamma}}
    \begin{pmatrix}
    \mathcal{M}[\Sigma^{\rFS}(\mu)](N) \\
    \mathcal{M}[f_{g}^{\rFS}(\mu)](N)
    \end{pmatrix}
\end{align}
for evolution matrix
\begin{align}
    \bm{\mathrm{\gamma}} =
    \begin{pmatrix}
        \gamma_{qq}(N) & \gamma_{qg}(N) \\
        \gamma_{gq}(N) & \gamma_{gg}(N)
    \end{pmatrix},
\end{align}
the evolution matrix $\bm{\mathrm{\gamma}}$ becomes approximately diagonal 
in the large-$N$ limit as the off-diagonal entries vanish,
and so the equations again decouple.
Equivalently, from the eigenvalues of the evolution matrix,
\begin{align}
    \lambda_{\pm}(N)
	{} & =
    \frac{1}{2}\left(\Tr \bm{\mathrm{\gamma}} \pm \sqrt{\Tr(\bm{\mathrm{\gamma}})^2 - 4 \det(\bm{\mathrm{\gamma}})} \bm\right) \\ & {}
    \sim
    \frac{1}{2} \left( \Tr \bm{\mathrm{\gamma}} \pm \lvert\gamma_{qq} - \gamma_{gg}\rvert \right) = \gamma_{qq}, \gamma_{gg},
\end{align}
i.e.\ in the large-$N$ limit, \cref{eq:dglapsol_ns} also holds for the singlet and gluon.

As outlined in \cref{sec:disc_zto0}, splitting functions in alternative factorisation schemes
 do not necessarily have the same asymptotic behaviour as in the \msbar scheme,
 with the factorisation scheme transformation changing the leading behaviour of both small-$N$
 eigenvalues $\lambda_{\pm}$.
 
\begin{figure*}[tp]
\centering
\begin{subfigure}[t]{0.45\textwidth}
    \includegraphics[width=\textwidth]{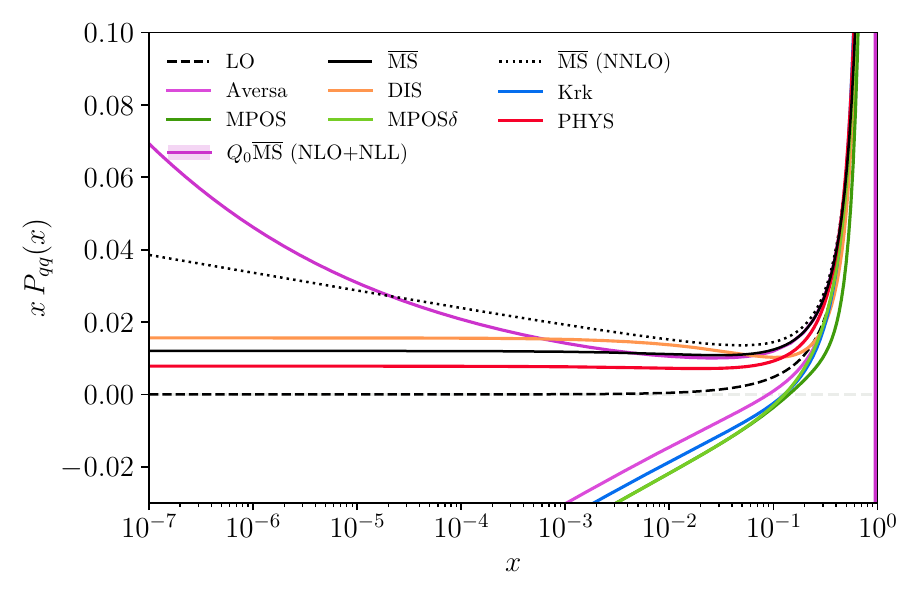}
\caption{$P_{qq}^{\rFS}$\label{fig:xspacePqqHELL}}
\end{subfigure}
\begin{subfigure}[t]{0.45\textwidth}
    \includegraphics[width=\textwidth]{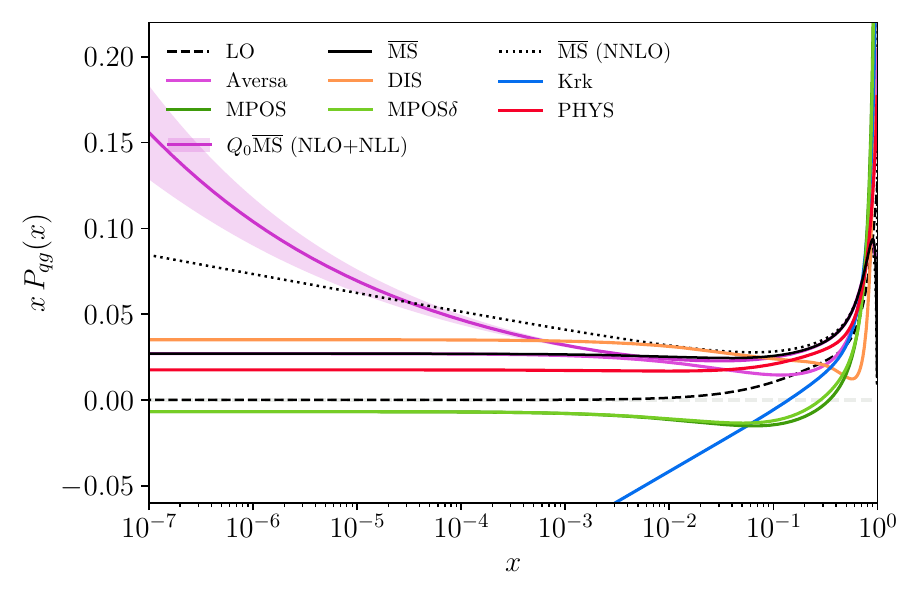}
\caption{$P_{qg}^{\rFS}$\label{fig:xspacePqgHELL}}
\end{subfigure}
\\
\begin{subfigure}[t]{0.45\textwidth}
    \includegraphics[width=\textwidth]{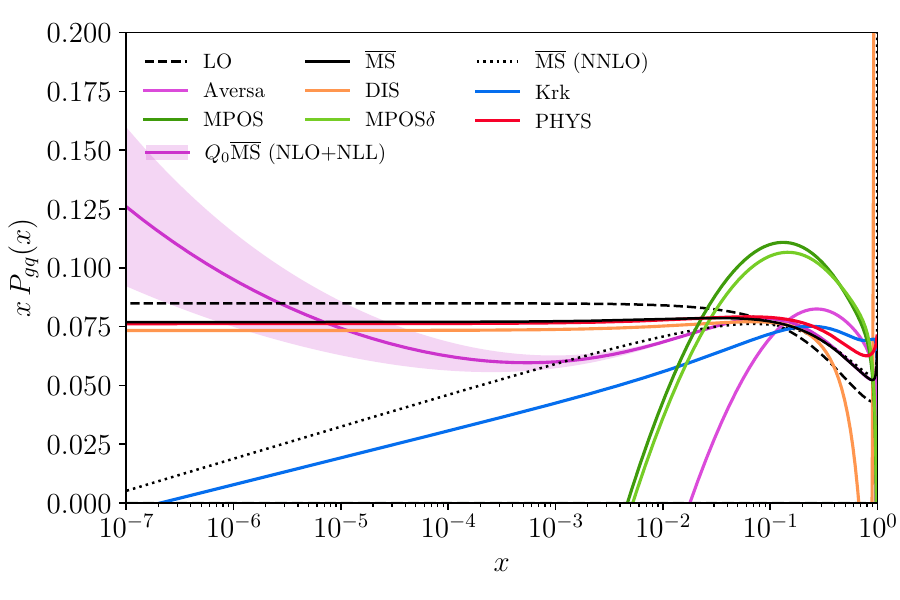}
\caption{$P_{gq}^{\rFS}$\label{fig:xspacePgqHELL}}
\end{subfigure}
\begin{subfigure}[t]{0.45\textwidth}
    \includegraphics[width=\textwidth]{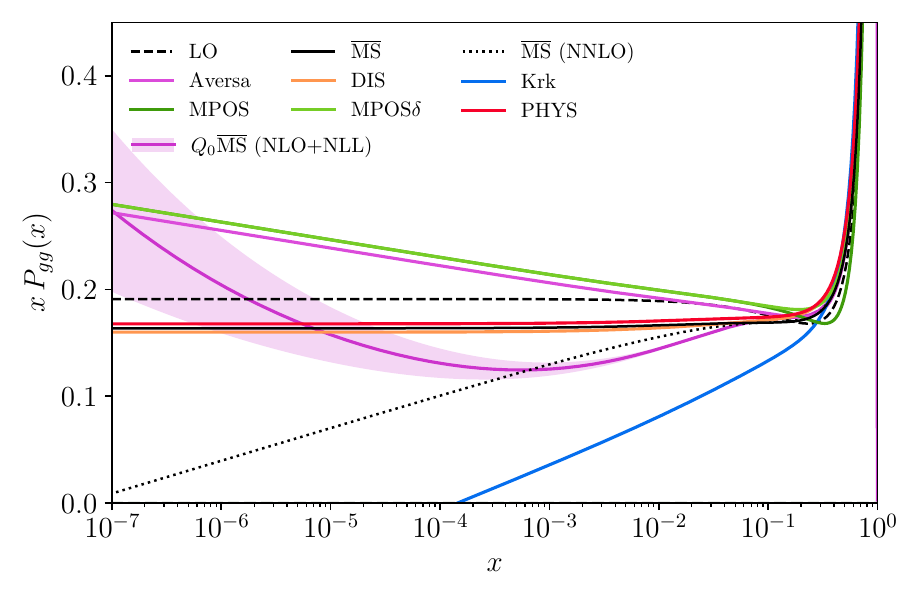}
\caption{$P_{gg}^{\rFS}$\label{fig:xspacePggHELL}}
\end{subfigure}
\caption{Splitting functions for the singlet sector, 
multiplied by $x$ (calculated with $\alphas = 0.2$, $\nf = 4$).
The NLO+NLL small-$x$ resummed results are calculated using HELL 4.0 \cite{Bonvini:2026cxp,Bonvini:emails}.
At this order, for $\rP_{qq}$ and $\rP_{qg}$ the $Q_0 \msbar$ scheme
is identical to the $\msbar$ scheme.}
\label{fig:xspacePHELL}
\end{figure*}

We do not consider the resummation of other schemes in detail here, but
in \cref{fig:xspacePHELL} we plot the small-$x$ behaviour of the NLO splitting functions in the schemes
considered here, alongside the \msbar NNLO splitting functions,
and NLO+NLL resummed splitting functions 
calculated using HELL 4.0 \cite{Bonvini:2026cxp,Bonvini:emails}.

As can be seen from the figures, while the asymptotic behaviour of the schemes as $x \to 0$ is
generally quite different from the resummed results,
over the range $x \gtrsim 10^{-5}$
the resummed splitting functions remain within the envelope of the named schemes,
and the scheme variation is of comparable magnitude to both the NNLO \msbar
correction and the NLL resummation effect.
Since evolution with these is routine, we expect that no numerical pathology
will arise for evolution in this region.
However, even in the absence of numerical pathologies, the evolved PDFs are expected to differ
qualitatively between the schemes, especially at small-$x$.

We emphasise that the physically-meaningful context in which to compare the schemes, including with higher-order calculations and
with resummation,
is on the level of their predictions for physical observables,
which we defer to future work.

\section{Conclusions}
\label{sec:conclusions}

In this work we have calculated the NLO DGLAP splitting functions in factorisation schemes other than the long-established \msbar and \dis schemes,
including 
the \krk and \phys schemes of direct phenomenological interest, and
a flexibly-parametrised `generalised' scheme (\gen)
introduced in \cite{Delorme:2025teo}.

These splitting functions are a necessary component of the generalisation of perturbative
QCD beyond the \msbar scheme, and a step towards being able to perform DGLAP evolution
and PDF fitting in alternative schemes.
This is a prerequisite for their application to QCD phenomenology,
and for the understanding of factorisation-scheme uncertainty at NLO and beyond.

We find that the familiar behaviour of the \msbar scheme
in both limits relevant to the resummation programme
is atypical among this family of schemes:
in the large-$x$ limit, its degree of singularity is limited to
the (in Mellin space, single-logarithmic) $\mathcal{D}_0$
distribution, to all perturbative orders;
in the small-$x$ limit, long-understood `accidental' cancellations
of the coefficient of the leading-logarithm render the \msbar
scheme less divergent at NLO than expected in general.

In the former case, large-$x$, this uncharacteristic behaviour can be expected
to extend to, and indeed, be exacerbated at, higher orders; the asymptotic large-$x$ behaviour
in a general factorisation scheme may be expected to contain
plus-distributions up to the order that arises in a perturbative
calculation of the same order, i.e.\ schematically, $\mathcal{D}_{2k-1}$ for an N$^k$LO calculation.
In the latter case, small-$x$, the coefficients in the \msbar scheme are only
sporadically vanishing, leading to an impression of perturbative convergence
that is not likely to generalise from the \msbar scheme to schemes in which
the logarithmic coefficients are non-vanishing.

Within the framework of the parametrised \gen scheme,
we have identified the coefficients that control the behaviour in each limit,
and identified those
relevant to the 
DGLAP evolution of the non-singlet distributions, as well as
to the eigendecomposition of the singlet sector in both
small- and large-$x$ limits.
We interpret the leading behaviour in each limit as corresponding to an effective
evolution scale, modifying the scale of the running of the strong coupling
from the central scale $\mu$ to a scale commensurate with the associated emissions.
This interpretation suggests that it is possible to design new schemes with specific effective scales missing from our set of named schemes, such as $\mu_{\mathrm{eff}} = \mu \sqrt{ (1-x) / x }$,
and that the dominant effect of scheme-variation at LO can be captured, and higher-order terms resummed,
by using a modified $x$-dependent effective scale for the running coupling.

Among the structural properties of the generalised scheme,
we note that momentum conservation can be imposed to reduce the dimensionality of the parameter space (as in \cite{Delorme:2025teo}).
This dimensionality can further be reduced by imposing
Casimir scaling of the anomalous dimensions
and the consistent treatment of soft gluon and high-energy gluon contributions
between channels.

Taken together, our results suggest that a systematic treatment of factorisation-scheme variation and the associated scheme uncertainty,
extending beyond specific named schemes,
may be feasible
within a reduced, physically-interpretable space of alternative schemes.
This opens the door to systematic scheme-variation studies at NLO and beyond.

Finally, we note that, while our focus is on the initial-state (`spacelike') evolution of parton distribution functions,
the question of factorisation scheme dependence applies equally to the final-state `timelike' evolution of fragmentation functions, and our results can be converted
to timelike evolution kernels by the addition of the known \msbar conversion term.

\section*{Acknowledgements}
The authors wish to thank 
Ted Rogers and Valerio Bertone
for correspondence on subjects related to those discussed here,
and Ted Rogers for comments on the manuscript.
We are grateful to Simone Marzani and Marco Bonvini for correspondence about the results of \cite{Bonvini:2016wki,Bonvini:2017ogt,Bonvini:2018xvt,Bonvini:2026cxp},
and to Marco Bonvini for providing the HELL 4.0 results used in \cref{fig:xspacePHELL}.

The work of A.S.\ and A.K.\ is supported by the National Science Centre, Poland,
under OPUS grant No.\ 2025/57/B/ST2/04034.
A.S.\ is also supported by the Priority Research Area Digiworld
under the program `Excellence Initiative -- Research University'
at the Jagiellonian University in Kraków.
S.D.\ is grateful for the support of the National Science Centre, Poland under MAESTRO grant No.\ 2023/50/A/ST2/00224.
We gratefully acknowledge the Polish high-performance computing infrastructure PLGrid (HPC Centre: ACK Cyfronet AGH) for providing computing facilities and support within computational grant PLG/2025/018371.

\clearpage
\appendix
\settocdepth{section}
\crefalias{section}{appendix}

\section{Notation and conventions}
\label{sec:app_conventions}

Perturbative expansions are consistently indexed according to the convention
\begin{align}
F(\alphas, \mu)
=
F^{(0)}(\mu) + \left(\frac{\alphas(\mu)}{2\pi}\right) F^{(1)}(\mu) + \dots 
= 
\sum_{k} 
\left(\frac{\alphas(\mu)}{2\pi}\right)^k 
F^{(k)}(\mu).
\label{eq:pertexpconv}
\end{align}
We expand the matrix of scheme-transformation convolution kernels %
of \cref{eq:fstransformations_faFSvec,eq:fstransformations_faFS} perturbatively as
\begin{align}
	\label{eq:fstransformations_pertexp}
	\mathbb{K}^{\msbar \to \rFS}_{ab} \left(z, \mu\right)
	= {} &
	\delta_{ab} \; \delta(1-z)
	+
	\frac{\alphas(\mu)}{2\pi}
	\;
	\rK^{\msbar \to \rFS}_{ab}(z, \mu) + \order{\alphas^2}.
\end{align}

The QCD $\beta$-function,
\begin{align}
  \label{eq:betafunction}
  \mu^2 \frac{\partial}{\partial \mu^2} \alphas(\mu)
  &=
  \beta(\alphas (\mu))
\end{align}
therefore has the perturbative expansion
\begin{align}
    \beta^{(0)} & {} = 0 &
    \beta^{(1)} & {} = 0 \\
    \beta^{(2)} & {} = \frac{\pi}{3} \left( 4 \nf \tR - 11 \ca \right) &
    \beta^{(3)} & {} = \frac{\pi}{3} \left(10\ca\nf\tr + 6\cf\nf\tr -17\ca^2 \right).
\end{align}

The leading-order colour-factor-stripped DGLAP splitting functions are given in four dimensions by
\begin{align}
	\pqq &= \frac{1+z^2}{1-z}	  %
	&
	\pqg &= z^2 + (1-z)^2 
	\\
	\pgq &= \frac{1 + (1-z)^2}{z} 
	&
	\pgg &= %
	  \frac{1}{1-z} + \frac{1}{z} - 2 + z(1-z)
\end{align}
and the leading-order expansion of the DGLAP kernels, in all schemes,%
\footnote{As discussed in \cref{sec:facSchemes}, scheme-independence here holds subject to $\partial \rK_{ab}^{\msbar \to \rFS(1)} (\mu)  / \partial \mu^2 = 0$.} is
\begin{align}
\label{eq:LOkernels}
    \rP^{\rFS(1)}_{qq} (z) &= \cf \, \left[ \pqq \right]_+
    &
    \rP^{\rFS(1)}_{q_ig} (z) &= \tr \; \pqg
    \\
    \rP^{\rFS(1)}_{gq} (z) &= \cf \; \pgq
    &
    \rP^{\rFS(1)}_{gg} (z) &= 2 \ca \; \left(
    \left[\frac{1}{1-z}\right]_+ + \frac{1}{z} - 2 + z(1-z)
    \right)
    - \frac{\beta^{(2)}}{2\pi} \delta(1-z).
\end{align}
The colour factors $C_{ab}$ used in \cref{eq:effscaleexp}
may be read off as $C_{qq,gq} = \cf$ for $\rP_{qq}$, $\rP_{gq}$;
$C_{q_ig}=\tr$ for $\rP_{q_ig}$, and $C_{gg}=2 \ca$ for $\rP_{gg}$.

The Mellin moments of an $x$-space function $f$ are defined as
\begin{align}
    \mathcal{M}[f](N)
    =
    \int_0^1 z^{N-1} \, f(z) \; \dd z,
\end{align}
and we define the anomalous dimensions as
\begin{align}
    \gamma_{ab}^{\rFS} (N)
    = 
    \mathcal{M}[\rP^{\rFS}_{ab}](N)
\end{align}
and their matrix for the evolution of the singlet system
\begin{align}
    \bm{\mathrm{\gamma}} (N) =
    \begin{pmatrix}
        \gamma_{qq}(N) & \gamma_{qg}(N) \\
        \gamma_{gq}(N) & \gamma_{gg}(N)
    \end{pmatrix}.
\end{align}

The Mellin transforms used in \cref{sec:asymptotics}
are collected here for reference:
\begin{align}
    \mathcal{M}[\mathcal{D}_0(z)](N) &= - H_{N-1} 
    \underset{N \to \infty}{\sim} - \log N - \egamma + \order{N^{-1}}
    \\
    \mathcal{M}[\mathcal{D}_1(z)](N) &= 
    \frac{1}{2}\left( H_{N-1}^2 + H_{N-1}^{(2)} \right)
    \underset{N \to \infty}{\sim} \frac{1}{2} \log^2 N + \egamma \log N + \frac{1}{2} \egamma^2 + \frac{\pi^2}{12} + \order{N^{-1} \log N}
    \\
    \mathcal{M}[ z^k ](N) &= \frac{1}{N+k}
    \\
    \mathcal{M}[ z^k \log (1-z) ](N) &= - \frac{H_{N+k}}{N+k}
    \\
    \mathcal{M}[ z^{-1} \log^k z ](N) &= (-1)^{k} \frac{k!}{(N-1)^{k+1}}
    \\
    \mathcal{M}[ z^k \log z ](N) &= - \frac{1}{(N+k)^2}
    \\
    \mathcal{M}\left[z^k \frac{\log z}{1-z}\right](N) &=
    H_{N+k-1}^{(2)} - \frac{\pi^2}{6}
    \\
    \mathcal{M}[\delta(1-z)](N) &= 1.
\end{align}
Here $H_N^{(r)}$ denotes the $N$\textsuperscript{th} generalised harmonic number of order $r$,
\begin{align}
    H_N^{(r)} = \sum_{k=1}^{N} \frac{1}{k^{r}};
\end{align}
additionally, we omit the order for the harmonic numbers of order 1: $H_N^{(1)}\equiv H_N$.
The asymptotic expansions of $\mathcal{M}[\mathcal{D}_{0,1}]$ follow from
\begin{align}
    H_N = \log N + \gamma_{\mathrm{E}} + \mathcal{O}\left(\frac{1}{N}\right),
\end{align}
where $\egamma \approx 0.577$ is the Euler--Mascheroni constant.

\section{DIS-scheme splitting functions}
\label{sec:appMSbarDIS}

\setcounter{equation}{0}

In this appendix we present the NLO splitting functions
in the \dis scheme, in compact notation consistent with the other schemes in \cref{sec:calcres}.
To our knowledge, these have not previously been published in closed-form.

\input{splitfns/PDIS_formatted.tex}

\newpage

\bibliographystyle{JHEP}
\bibliography{refs.bib}

\end{document}

%% file: splitfns/PKrk_formatted.tex
\begin{align}
\label{eq:PKrkqiqi}
P_{q_iq_i}^{\text{Krk}\,(2)}(x) &= C_F n_f T_R \Big[
 \begin{aligned}[t]
& \frac{8}{3} \left[\frac{\log (1-x)}{1-x}\right]_+ 
  -\frac{20}{9} \left[\frac{1}{1-x}\right]_+ 
  -\frac{4}{3} \log (x) p_{qq}(x) \\
  & -\frac{4}{3} (1 + x) \log (1-x)
  +\frac{4}{9} (1 + 4 x) 
  -\left(3 + \frac{4 \pi ^2}{9}\right) \delta (1-x) \Big]
\end{aligned} \\ \notag
 &\quad + C_F T_R \Big[
 \begin{aligned}[t]
& -(x + 1) \log ^2(x)
  +\left(\frac{4 x^2}{3}+2 x+\frac{4}{3 x}\right) \log (x)
  -\frac{26 x^2}{9} +5 x +\frac{26}{9 x} -5 \Big]
\end{aligned} \\ \notag
 &\quad + C_F^2 \Big[
 \begin{aligned}[t]
& -2 \log (1-x) \log (x) p_{qq}(x) - \frac{1}{2} (x + 1) \log ^2(x) \\
  & +\frac{\left(-2 x^2+2 x+3\right)}{x-1} \log (x)
  +5 (x-1)
  +\left(6 \zeta_3+\frac{3}{8}-\frac{\pi ^2}{2}\right) \delta (1-x) \Big]
\end{aligned} \\ \notag
 &\quad + C_A C_F \Big[
 \begin{aligned}[t]
& -\frac{22}{3} \left[\frac{\log (1-x)}{1-x}\right]_+
  +\left(\frac{67}{9}-\frac{\pi ^2}{3}\right) \left[\frac{1}{1-x}\right]_+
  +\frac{1}{2} \log ^2(x) p_{qq}(x) +\frac{1}{6} \pi ^2 (1 + x) \\
  & +\frac{\left(8 x^2+14\right)}{3-3 x} \log (x) +\frac{11}{3} (1 + x) \log (1-x)
  +\frac{1}{9} (10-77 x)
  +\left(-3 \zeta_3+\frac{17}{2}+\frac{11 \pi ^2}{9}\right) \delta (1-x) \Big]
\end{aligned}
\end{align}

\begin{align}
\label{eq:PKrkqiqk}
P_{q_iq_k}^{\text{Krk}\,(2)}(x) = P_{q_i\overline{q}_k}^{\text{Krk}\,(2)}(x) &= C_F T_R \Big[
 \begin{aligned}[t]
& -(x + 1) \log ^2(x)
  +\left(\frac{4 x^2}{3}+2 x+\frac{4}{3 x}\right) \log (x)
  -\frac{26 x^2}{9} +5 x +\frac{26}{9 x} -5 \Big]
\end{aligned}
\end{align}

\begin{align}
\label{eq:PKrkqiqbi}
P_{q_i\overline{q}_i}^{\text{Krk}\,(2)}(x) &= C_F T_R \Big[
 \begin{aligned}[t]
& -(x + 1) \log ^2(x) 
  +\left(\frac{4 x^2}{3}+2 x+\frac{4}{3 x}\right) \log (x)
  -\frac{26 x^2}{9} +5 x +\frac{26}{9 x} -5 \Big]
\end{aligned} \\ \notag
 &\quad + C_F(C_A - 2 C_F) \Big[
 \begin{aligned}[t]
& 2 \text{Li}_2(-x) p_{qq}(-x) +\frac{1}{6} \pi ^2 p_{qq}(-x) -\frac{1}{2} \log ^2(x) p_{qq}(-x) \\
  & +2 \log (1 + x) \log (x) p_{qq}(-x) 
  -(x + 1) \log (x) 
  +2 (x-1) \Big]
\end{aligned}
\end{align}

\begin{align}
\label{eq:PKrkqig}
P_{q_ig}^{\text{Krk}\,(2)}(x) &= n_f T_R^2 \Big[
 \begin{aligned}[t]
& -\frac{59}{36} p_{qg}(x) \Big]
\end{aligned} \\ \notag
 &\quad + C_F T_R \Big[
 \begin{aligned}[t]
& -\log ^2(1-x) p_{qg}(x) +\left(2 x^2-x+\frac{1}{2}\right) \log ^2(x) \\
  & -3 \log (1-x) p_{qg}(x) 
  +2 (1-x) (1-3x) \log (x)
  +\frac{9 x^2}{2} -8 x +\frac{7}{4} \Big]
\end{aligned} \\ \notag
 &\quad + C_A T_R \Big[
 \begin{aligned}[t]
& -2 \text{Li}_2(-x) p_{qg}(-x) +\log ^2(1-x) p_{qg}(x) -2 \log (x) \log (1-x) p_{qg}(x) \\
  & -2 \log (x) \log (1 + x) p_{qg}(-x) -\frac{1}{3} \pi ^2 \left(2 x^2 + 1\right) -(2 x + 1) \log ^2(x) \\
  & +\left(\frac{31 x^2}{3}+4 x+\frac{4}{3 x}\right) \log (x)
  +\frac{-390 x^3+686 x^2-91 x+208}{72 x} \Big]
\end{aligned}
\end{align}

\begin{align}
\label{eq:PKrkgq}
P_{gq}^{\text{Krk}\,(2)}(x) &= C_F n_f T_R \Big[
 \begin{aligned}[t]
& \frac{4}{3} \log (1-x) p_{gq}(x) -\frac{4}{3} \log (x) p_{gq}(x)
  -\frac{7}{12} p_{gq}(x) \Big]
\end{aligned} \\ \notag
 &\quad + C_F^2 \Big[
 \begin{aligned}[t]
& -\frac{1}{3} \pi ^2 p_{gq}(x) +\log ^2(1-x) p_{gq}(x) -2 \log (x) \log (1-x) p_{gq}(x) +\frac{1}{2} (x-2) \log ^2(x)
  +\frac{5 x}{4} +\frac{11}{2 x} -7 \Big]
\end{aligned} \\ \notag
 &\quad + C_A C_F \Big[
 \begin{aligned}[t]
& -2 \text{Li}_2(-x) p_{gq}(-x) -\log ^2(1-x) p_{gq}(x) -2 \log (x) \log (1 + x) p_{gq}(-x) \\
  & +\frac{\left(x^2+2\right)}{3 x} \pi ^2 +(x+2) \log ^2(x)
  -\frac{11}{3} \log (1-x) p_{gq}(x) 
  \\ &
  +\frac{1}{3} \left(-4 x^2+5 x+\frac{9}{x}-46\right) \log (x)
  +\frac{208 x^3-93 x^2+690 x-538}{72 x} \Big]
\end{aligned}
\end{align}

\begin{align}
\label{eq:PKrkgg}
P_{gg}^{\text{Krk}\,(2)}(x) &= n_f^2 T_R^2 \Big[
 \begin{aligned}[t]
& \frac{59}{54} \delta (1-x) \Big]
\end{aligned} \\ \notag
 &\quad + C_F n_f T_R \Big[
 \begin{aligned}[t]
& -2 (1 + x) \log ^2(x) 
  +\left(\frac{8 x^2}{3}-4 x-\frac{8}{3 x}-4\right) \log (x) 
  +10 (x-1) 
  -\delta (1-x) \Big]
\end{aligned} \\ \notag
 &\quad + C_A n_f T_R \Big[
 \begin{aligned}[t]
& \frac{8}{3} \left[\frac{\log (1-x)}{1-x}\right]_+ 
  -\frac{20}{9} \left[\frac{1}{1-x}\right]_+ \\
  & -\frac{8 \left(x^3-x^2+2 x-1\right)}{3 x} \log (1-x) +\frac{4 \left(x^4-3 x^3+3 x^2-x+1\right)}{3 (x-1) x} \log (x) \\
  & +\frac{2 \left(23 x^3-19 x^2+29 x-23\right)}{9 x}
  -\left(\frac{1619}{216} + \frac{2 \pi ^2}{9}\right) \delta (1-x) \Big]
\end{aligned} \\ \notag
 &\quad + C_A^2 \Big[
 \begin{aligned}[t]
& -\frac{22}{3} \left[\frac{\log (1-x)}{1-x}\right]_+ 
  +\left(\frac{67}{9}-\frac{\pi ^2}{3}\right) \left[\frac{1}{1-x}\right]_+  \\
  & -4 \text{Li}_2(-x) p_{gg}(-x) -4 \log (1 + x) \log (x) p_{gg}(-x) -4 \log (1-x) \log (x) p_{gg}(x) \\
  & -\frac{2 \left(-x^2+x+1\right)^2}{x^2-1} \log ^2(x) +\frac{\left(2 x^3+2 x^2+4 x+3\right)}{3 x+3} \pi ^2
  +\frac{22 \left(x^3-x^2+2 x-1\right)}{3 x} \log (1-x) \\
  & +\frac{\left(55 x^4-77 x^3+69 x^2-47 x+11\right)}{3 x-3 x^2} \log (x)
  -\frac{1}{18} (109 x + 25)
  +\left(3 \zeta_3+\frac{4903}{432}+\frac{11 \pi ^2}{18}\right) \delta (1-x) \Big]
\end{aligned}
\end{align}

%% file: splitfns/PPHYS_formatted.tex
\begin{align}
\label{eq:PPHYSqiqi}
P_{q_iq_i}^{\text{PHYS}\,(2)}(x) &= C_F n_f T_R \Big[
 \begin{aligned}[t]
& \frac{4}{3} \left[\frac{\log (1-x)}{1-x}\right]_+
  -\frac{20}{9} \left[\frac{1}{1-x}\right]_+
  -\frac{2}{3} \log (x) p_{qq}(x) -\frac{2}{3} (1 + x) \log (1-x) \\
  & +\frac{4}{9} (1 + 4 x)
  -\left(2 + \frac{2 \pi ^2}{9}\right) \delta (1-x) \Big]
\end{aligned} \\ \notag
 &\quad + C_F T_R \Big[
 \begin{aligned}[t]
& -(x + 1) \log ^2(x)
  +\left(\frac{4 x^2}{3}-1\right) \log (x)
  -\frac{13 x^2}{9} +5 x +\frac{13}{9 x} -5 \Big]
\end{aligned} \\ \notag
 &\quad + C_F^2 \Big[
 \begin{aligned}[t]
& -2 \log (1-x) \log (x) p_{qq}(x) -  \frac{1}{2} (x + 1) \log ^2(x) \\
  & +\frac{\left(-2 x^2+2 x+3\right)}{x-1} \log (x)
  +5 (x-1)
  +\left(6 \zeta_3+\frac{3}{8}-\frac{\pi ^2}{2}\right) \delta (1-x) \Big]
\end{aligned} \\ \notag
 &\quad + C_A C_F \Big[
 \begin{aligned}[t]
& -\frac{11}{3} \left[\frac{\log (1-x)}{1-x}\right]_+
  +\left(\frac{67}{9}-\frac{\pi ^2}{3}\right) \left[\frac{1}{1-x}\right]_+ 
  +\frac{1}{2} \log ^2(x) p_{qq}(x) +\frac{1}{6} \pi ^2 (1 + x) \\
  & +\frac{\left(5 x^2+17\right)}{6-6 x} \log (x) +\frac{11}{6} (1 + x) \log (1-x)
  +\frac{1}{9} (10-77 x)
  +\left(-3 \zeta_3+\frac{23}{4}+\frac{11 \pi ^2}{18}\right) \delta (1-x) \Big]
\end{aligned}
\end{align}

\begin{align}
\label{eq:PPHYSqiqk}
P_{q_iq_k}^{\text{PHYS}\,(2)}(x) = P_{q_i\overline{q}_k}^{\text{PHYS}\,(2)}(x) &= C_F T_R \Big[
 \begin{aligned}[t]
& -(x + 1) \log ^2(x)
  +\left(\frac{4 x^2}{3}-1\right) \log (x)
  -\frac{13 x^2}{9} +5 x +\frac{13}{9 x} -5 \Big]
\end{aligned}
\end{align}

\begin{align}
\label{eq:PPHYSqiqbi}
P_{q_i\overline{q}_i}^{\text{PHYS}\,(2)}(x) &= C_F T_R \Big[
 \begin{aligned}[t]
& -(x + 1) \log ^2(x)
  +\left(\frac{4 x^2}{3}-1\right) \log (x)
  -\frac{13 x^2}{9} +5 x +\frac{13}{9 x} -5 \Big]
\end{aligned} \\ \notag
 &\quad + C_F (C_A - 2 C_F) \Big[
 \begin{aligned}[t]
& 2 \text{Li}_2(-x) p_{qq}(-x) +\frac{1}{6} \pi ^2 p_{qq}(-x) -\frac{1}{2} \log ^2(x) p_{qq}(-x) \\
  & +2 \log (1 + x) \log (x) p_{qq}(-x)
  -(x + 1) \log (x)
  +2 (x-1) \Big]
\end{aligned}
\end{align}

\begin{align}
\label{eq:PPHYSqig}
P_{q_ig}^{\text{PHYS}\,(2)}(x) &= n_f T_R^2 \Big[
 \begin{aligned}[t]
& -\frac{29}{36} p_{qg}(x) \Big]
\end{aligned} \\ \notag
 &\quad + C_F T_R \Big[
 \begin{aligned}[t]
& 2 \text{Li}_2(x) p_{qg}(x) -\frac{1}{3} \pi ^2 p_{qg}(x) +\left(2 x^2-x+\frac{1}{2}\right) \log ^2(x) \\
  & -\frac{3}{2} \log (1-x) p_{qg}(x) +\left(3 x^2-3 x+\frac{1}{2}\right) \log (x)
  +4 x^2 -6 x +\frac{7}{4} \Big]
\end{aligned} \\ \notag
 &\quad + C_A T_R \Big[
 \begin{aligned}[t]
& -\text{Li}_2\left(x^2\right) p_{qg}(-x) -2 \log (1 + x) \log (x) p_{qg}(-x) -2 \log (1-x) \log (x) p_{qg}(x) \\
  & +8 x \text{Li}_2(x) -\frac{2 \pi ^2 x}{3} -(2 x + 1) \log ^2(x)
  +\left(\frac{31 x^2}{3}-1\right) \log (x)
  +\frac{-310 x^3+602 x^2-121 x+104}{72 x} \Big]
\end{aligned}
\end{align}

\begin{align}
\label{eq:PPHYSgq}
P_{gq}^{\text{PHYS}\,(2)}(x) &= C_F n_f T_R \Big[
 \begin{aligned}[t]
& -\frac{17}{12} p_{gq}(x) \Big]
\end{aligned} \\ \notag
 &\quad + C_F^2 \Big[
 \begin{aligned}[t]
& -2 \text{Li}_2(x) p_{gq}(x) -2 \log (1-x) \log (x) p_{gq}(x) +\frac{1}{2} (x-2) \log ^2(x) \\
  & -\frac{3}{2} \log (1-x) p_{gq}(x) +\left(\frac{3 x}{2}-2\right) \log (x)
  +\frac{x}{4} +\frac{1}{x} -3 \Big]
\end{aligned} \\ \notag
 &\quad + C_A C_F \Big[
 \begin{aligned}[t]
& -\text{Li}_2\left(x^2\right) p_{gq}(-x) -2 \log (1 + x) \log (x) p_{gq}(-x) -8 \text{Li}_2(x) +(x+2) \log ^2(x) \\
  & +\frac{2 \pi ^2}{3}
  -\frac{4}{3} \left(x^2+6\right) \log (x)
  +\frac{1}{72} \left(104 x^2+9 x-\frac{50}{x}+342\right) \Big]
\end{aligned}
\end{align}

\begin{align}
\label{eq:PPHYSgg}
P_{gg}^{\text{PHYS}\,(2)}(x) &= n_f^2 T_R^2 \Big[
 \begin{aligned}[t]
& \frac{29}{54} \delta (1-x) \Big]
\end{aligned} \\ \notag
 &\quad + C_F n_f T_R \Big[
 \begin{aligned}[t]
& -2 (1 + x) \log ^2(x)
  +\left(\frac{8 x^2}{3}-2\right) \log (x)
  -\frac{26 x^2}{9} +10 x +\frac{26}{9 x} -10
  -\delta (1-x) \Big]
\end{aligned} \\ \notag
 &\quad + C_A n_f T_R \Big[
 \begin{aligned}[t]
& \frac{4}{3} \left[\frac{\log (1-x)}{1-x}\right]_+ 
  -\frac{20}{9} \left[\frac{1}{1-x}\right]_+
  -\frac{4 \left(x^3-x^2+2 x-1\right)}{3 x} \log (1-x) -\frac{4}{3} (1 + x) \log (x) \\
  & +\frac{2 \left(23 x^3-19 x^2+29 x-23\right)}{9 x}
  -\frac{1013}{216} \delta (1-x) \Big]
\end{aligned} \\ \notag
 &\quad + C_A^2 \Big[
 \begin{aligned}[t]
& -\frac{11}{3} \left[\frac{\log (1-x)}{1-x}\right]_+
  +\left(\frac{67}{9}-\frac{\pi ^2}{3}\right) \left[\frac{1}{1-x}\right]_+ \\
  & -4 \text{Li}_2(-x) p_{gg}(-x) -4 \log (1 + x) \log (x) p_{gg}(-x) -4 \log (1-x) \log (x) p_{gg}(x) \\
  & -\frac{2 \left(-x^2+x+1\right)^2}{x^2-1} \log ^2(x) +\frac{ \left(2 x^3+2 x^2+4 x+3\right)}{3 x+3} \pi ^2
  +\frac{1}{3} \left(-44 x^2+11 x-25\right) \log (x) 
  \\ &
  +\frac{11 \left(x^3-x^2+2 x-1\right)}{3 x} \log (1-x)
  -\frac{1}{18} (109 x + 25)
  +\left(3 \zeta_3+\frac{3385}{432}\right) \delta (1-x) \Big]
\end{aligned}
\end{align}

%% file: splitfns/PGEN_formatted.tex
\begin{align}
\label{eq:PGENqiqi}
P_{q_iq_i}^{\text{GEN}\,(2)}(x) &= C_F n_f T_R \Big[
 \begin{aligned}[t]
& \frac{4}{3} a_{qq} \left[\frac{\log (1-x)}{1-x}\right]_+
  -\frac{2}{9} \left(3 b_{qq}+10\right) \left[\frac{1}{1-x}\right]_+ 
  -\frac{2}{3} \left(c_{qq}+1\right) \log (x) p_{qq}(x) \\
  & -\frac{2}{3} (1 + x) a_{qq} \log (1-x) 
  +\frac{2}{9} (11 x-1)
  -\left(\frac{2}{3} \Delta_{qq}+\frac{1}{6}+\frac{2}{9} \pi ^2\right) \delta (1-x) \Big]
\end{aligned} \\ \notag
 &\quad + C_F T_R \Big[
 \begin{aligned}[t]
& 2 (1 + x) \text{Li}_2(x) \left(a_{gq}-a_{qg}\right) -\frac{1}{3} \pi ^2 (1 + x) \left(a_{gq}-a_{qg}\right)
  +(1 + x) \log ^2(x) \left(c_{gq}-c_{qg}-1\right) \\
  & +\log (x) \left(
  -\frac{2}{3} \left(2 x^2+3 x+3\right) a_{gq}+x a_{qg}
  +c_{gq} \left(x +\frac{4}{3 x}+2\right)
  +c_{qg} \left(\frac{4 x^2}{3}+2 x+1\right)
  +\frac{8 x^2}{3}+5 x+1
  \right) \\
  & +\left(a_{qg}-a_{gq}\right) \frac{(1-x) (4x^2 + 7x + 4)}{3 x}
  \log (1-x) \\
  & +\frac{(x-1) \left(3 x a_{gq}+\left(13 x^2+10 x+13\right) a_{qg}
  -13 (c_{gq}+c_{qg})\left(x^2 + x + 1\right)
  -56 x^2-2 x-20\right)}{9 x} \Big]
\end{aligned} \\ \notag
 &\quad + C_F^2 \Big[
 \begin{aligned}[t]
& -2 \log (1-x) \log (x) p_{qq}(x) -\frac{1}{2} (x + 1) \log ^2(x)
  +\frac{\left(-2 x^2+2 x+3\right)}{x-1} \log (x) \\
  & +5 (x-1)
  +\left(6 \zeta_3+\frac{3}{8}-\frac{\pi ^2}{2}\right) \delta (1-x) \Big]
\end{aligned} \\ \notag
 &\quad + C_A C_F \Big[
 \begin{aligned}[t]
& -\frac{11}{3} a_{qq} \left[\frac{\log (1-x)}{1-x}\right]_+
  +\left(\frac{11}{6} b_{qq}-\frac{\pi ^2}{3}+\frac{67}{9}\right) \left[\frac{1}{1-x}\right]_+
  +\frac{1}{2} \log ^2(x) p_{qq}(x) +\frac{1}{6} \pi ^2 (1 + x) \\
  & +\frac{11}{6} (1 + x) a_{qq} \log (1-x) -\frac{11 \left(x^2+1\right) c_{qq}+5 x^2+17}{6 (x-1)} \log (x)
  +\frac{1}{18} (53-187 x) \\
  & +\left(\frac{11}{6} \Delta_{qq}-3 \zeta_3+\frac{11 \pi ^2}{18}+\frac{17}{24}\right) \delta (1-x) \Big]
\end{aligned}
\end{align}

\begin{align}
\label{eq:PGENqiqk}
P_{q_iq_k}^{\text{GEN}\,(2)}(x) =
P_{q_i\overline{q}_k}^{\text{GEN}\,(2)}(x)
&= C_F T_R \Big[
 \begin{aligned}[t]
& 2 (1 + x) \text{Li}_2(x) \left(a_{gq}-a_{qg}\right) -\frac{1}{3} \pi ^2 (1 + x) \left(a_{gq}-a_{qg}\right)
  +(1 + x) \log ^2(x) \left(c_{gq}-c_{qg}-1\right) \\
  & +\log (x) \biggl(-\frac{2}{3} \left(2 x^2+3 x+3\right) a_{gq}+x a_{qg}
  +c_{gq} \left( x +\frac{4}{3 x}+2 \right) \\
  &\qquad \qquad
  +  c_{qg} \left( \frac{4 x^2}{3}+2 x+1 \right)
  +\frac{8 x^2}{3}+5 x+1 \biggr) \\
  & +\left(a_{qg}-a_{gq}\right) \frac{(1-x) (4x^2 + 7x + 4)}{3 x}
  \log (1-x) \\
  & +\frac{x-1}{9x}\left(3 x a_{gq}+\left(13 x^2+10 x+13\right) a_{qg}
  - 13 (c_{gq} + c_{qg}) (x^2 + x + 1)
  -56 x^2-2 x-20\right) \Big]
\end{aligned}
\end{align}

\begin{align}
\label{eq:PGENqiqbi}
P_{q_i\overline{q}_i}^{\text{GEN}\,(2)}(x) &= C_F T_R \Big[
 \begin{aligned}[t]
& 2 (1 + x) \text{Li}_2(x) \left(a_{gq}-a_{qg}\right) -\frac{1}{3} \pi ^2 (1 + x) \left(a_{gq}-a_{qg}\right)
  +(1 + x) \log ^2(x) \left(c_{gq}-c_{qg}-1\right) \\
  & +\log (x) \left(-\frac{2}{3} \left(2 x^2+3 x+3\right) a_{gq}+x a_{qg}
  +c_{gq} \left(x +\frac{4}{3 x}+2\right)
  +c_{qg} \left(\frac{4 x^2}{3}+2 x +1\right)
  +\frac{8 x^2}{3}+5 x+1\right) \\
  & +\left(a_{qg}-a_{gq}\right) \frac{(1-x) (4x^2 + 7x + 4)}{3 x}
  \log (1-x) \\
  & +\frac{(x-1) \left(3 x a_{gq}
  +\left(13 x^2+10 x+13\right) a_{qg}
  - 13 (c_{gq} + c_{qg}) (x^2 + x + 1)
  -56 x^2-2 x-20\right)}{9 x} \Big]
\end{aligned} \\ \notag
 &\quad + C_F (C_A - 2 C_F) \Big[
 \begin{aligned}[t]
& 2 \text{Li}_2(-x) p_{qq}(-x) +\frac{1}{6} \pi ^2 p_{qq}(-x) -\frac{1}{2} \log ^2(x) p_{qq}(-x) \\
  & +2 \log (1 + x) \log (x) p_{qq}(-x)
  -(x + 1) \log (x)
  +2 (x-1) \Big]
\end{aligned}
\end{align}

\begin{align}
\label{eq:PGENqig}
P_{q_ig}^{\text{GEN}\,(2)}(x) &= C_F T_R \Big[
 \begin{aligned}[t]
& 2 \log (x) \log (1-x) p_{qg}(x) \left(a_{qg}-c_{qq}-1\right) +\log ^2(1-x) \left(-2 a_{qg}+a_{qq}+1\right) p_{qg}(x) \\
  & +\text{Li}_2(x) \left((1-2 x) a_{qg}+\left(4 x^2-2 x+1\right) a_{qq}-2 p_{qg}(x) %
  \left(c_{qg}+c_{qq}\right)\right) \\
  & +\frac{1}{6} \pi ^2 \left(
  \left(4 x^2-2 x+1\right) (a_{qg} - a_{qq})
  +2 p_{qg}(x) %
  \left(c_{qg}+c_{qq}-1\right)\right) \\
  & -\frac{1}{2} \left(4 x^2-2 x+1\right) \log ^2(x) \left(c_{qg}-c_{qq}-1\right) \\
  & +\log (x) \left(x a_{qg}+(3 x-2) x a_{qq}
  + p_{qg}(x) b_{qq}
  -\left( x -\frac{1}{2} \right) c_{qg}
  + ( -3 x + 1 ) c_{qq}
  +4 x^2-2 x+\frac{3}{2}\right) \\
  & +\log (1-x) \left(\left(\frac{1}{2}-2 x\right) a_{qg}
  + (1-x) (3x-2) a_{qq}
  - p_{qg}(x) b_{qq}
  -4 x^2+4 x\right) \\
  & +\frac{1}{2} \biggl(
  (1-x)(1-5x)
  (-a_{qg} + c_{qg} + c_{qq})
  +2 (x-1)^2 a_{qq}
  +2( 1-x ) (1-3x) b_{qq}
  \\ & \qquad \qquad
  - 2 p_{qg}(x) \Delta_{qq}
  +20 x^2-29 x+14\biggr) \Big]
\end{aligned} \\ \notag
 &\quad + C_A T_R \Big[
 \begin{aligned}[t]
& \log (x) \log (1-x) p_{qg}(x) \left(2 c_{gg}-2 a_{qg}\right) +\log ^2(1-x) \left(-a_{gg}+2 a_{qg}-1\right) p_{qg}(x) \\
  & -2 \text{Li}_2(-x) p_{qg}(-x) -2 \log (x) \log (1 + x) p_{qg}(-x) \\
  & +2 \text{Li}_2(x) \left((1 + 4 x) a_{gg}-2 \left(x^2+x+1\right) a_{qg}+p_{qg}(x)%
  \left(c_{gg}+c_{qg}\right)\right) \\
  & +\frac{1}{3} \pi ^2 \left(
  (1 + 4 x) (a_{qg} - a_{gg})
  - p_{qg}(x) %
  (c_{gg} + c_{qg})
  -2 x\right) 
  +\log ^2(x) \left(
  (1 + 4 x) (c_{gg}-c_{qg})
  -2 x-1\right) \\
  & +\log (x) \biggl(-\frac{2}{3} \left(11 x^2+3 x+3\right) a_{gg}+(3 x+2) x a_{qg}
  -p_{qg}(x) %
  b_{gg}
  + \left(-3 x^2 +6 x +\frac{4}{3 x}+2 \right) c_{gg}
  \\ & \qquad \qquad
  + \left( \frac{22 x^2}{3}-2 x + 1 \right) c_{qg}
  +\frac{44 x^2}{3}+8 x+1\biggr) \\
  & +\log (1-x) \left( \frac{
  (1-x) (31x^2 + 7x + 4)
  }{3 x}
  (a_{qg} - a_{gg})
  + p_{qg}(x) %
  b_{gg}
  +4 (x-1) x \right) \\
  & + \frac{1}{18x} \biggl(
  3 x (1-x) (21 x - 5)
  a_{gg}
  - 2 (1-x) (58x^2 + 10x + 13) a_{qg}
  + 6 x (1-x) (9x-3)
  b_{gg}
  \\ & \qquad
  + (1-x)(71x^2 + 17x + 26)
  (c_{gg} + c_{qg})
  +18 x (2 x^2 - 2 x + 1) \Delta_{gg}
  -436 x^3+450 x^2-36 x+40 \biggr) \Big]
\end{aligned}
\end{align}

\begin{align}
\label{eq:PGENgq}
P_{gq}^{\text{GEN}\,(2)}(x) &= C_F n_f T_R \Big[
 \begin{aligned}[t]
& \frac{4}{3} \left(a_{gq}-1\right) \log (1-x) p_{gq}(x) -\frac{4}{3} c_{gq} \log (x) p_{gq}(x)
  -\frac{8 \left(4 x^2-5 x+5\right)}{9 x} \Big]
\end{aligned} \\ \notag
 &\quad + C_F^2 \Big[
 \begin{aligned}[t]
& \log (x) \log (1-x) p_{gq}(x) \left(2 c_{qq}-2 a_{gq}\right) +\log ^2(1-x) \left(2 a_{gq}-a_{qq}-1\right) p_{gq}(x) \\
  & -\frac{\left(x^2-2 x+4\right) a_{gq}+(x-2) x a_{qq}-2 \left(x^2-2 x+2\right) \left(c_{gq}+c_{qq}\right)}{x} \text{Li}_2(x) \\
  & +\frac{
  (x-2) x (a_{qq} - a_{gq})
  -2 \left(x^2-2 x+2\right) \left(c_{gq}+c_{qq}\right)}{6 x} \pi ^2
  +\frac{1}{2} (x-2) \log ^2(x) \left(c_{gq}-c_{qq}+1\right) \\
  & +\log (x) \left(
  (x-2) (a_{gq} - b_{qq})
  +x \left(-a_{qq}\right)
  + c_{gq} \left( -\frac{1}{2}x + 1\right)
  + c_{qq} (-x+3)
  +\frac{7 x}{2}+2\right) \\
  & +\log (1-x) \left(\frac{
  (1-x) (3-2x)
  a_{qq}
  +\left(x^2-2 x+2\right) b_{qq}-5 x^2+6 x-6}{x}-\frac{1}{2} (x-4) a_{gq}\right) \\
  & + \frac{
  (1-x)(3x-7)
  a_{qq}
  +2 x (1-x) b_{qq}
  + (c_{gq} + c_{qq}) 
  (1-x)(5-x)
  + 2 \Delta_{qq} (x^2-2x+2)
  -7 x^2-5 x}{2 x} \Big]
\end{aligned} 
\\ \notag
 &\quad + C_A C_F \Big[
 \begin{aligned}[t]
& 2 \log (x) \log (1-x) p_{gq}(x) \left(a_{gq}-c_{gg}-1\right) 
  +\log ^2(1-x) \left(a_{gg}-2 a_{gq}+1\right) p_{gq}(x) \\
  & -2 \text{Li}_2(-x) p_{gq}(-x) -2 \log (x) \log (1 + x) p_{gq}(-x) \\
  & +\frac{4 \left(x^2+x+1\right) a_{gg}
  -2 \left(x (x+4) a_{gq}
  +\left(x^2-2 x+2\right) \left(c_{gg}+c_{gq}\right)\right)}{x} \text{Li}_2(x) \\
  & +\frac{
  2 \left(x^2+x+1\right) (a_{gq} - a_{gg})
  + (c_{gg} + c_{gq}) (x^2 - 2x + 2)
  +2 x}{3 x} \pi ^2 \\
  & +\frac{
  2 \left(x^2+x+1\right) (c_{gg} - c_{gq})
  +x (x+2)}{x} \log ^2(x) \\
  & +\frac{1}{3} \log (x) \biggl(-3 (x+4) a_{gg}
  +\left(4 x^2+6 x+24\right) a_{gq}
  +3 b_{gg} (x - 2)
  \\ & \qquad \qquad \qquad
  + c_{gg} ( -4 x^2-6 x+\frac{9}{x}-18 )
  +8 c_{gq} ( x -2)
  -8 x^2-15 x-36\biggr) \\
  & + \log (1-x) \frac{1}{3x} \biggl(
  -(1-x)(4x^2+7x+31)
  a_{gg}
  -\left(4 x^3+14 x^2+2 x-9\right) a_{gq}
  \\ & \qquad \qquad \qquad \qquad
  -3 b_{gg} (x^2 - 2x +2)
  +17 x^2-22 x+22 \biggr)  \\
  & + \frac{1}{18x} \biggl(
  2 (1-x) (13 x^2 + x + 67)
  a_{gg}
  - 3 (1-x)(15+x)
  a_{gq}
  +18 x (x-1) b_{gg}
  \\ & \qquad \qquad
  - (c_{gg} + c_{gq}) (1-x) (26x^2 + 17 x + 71)
  - 18 \Delta_{gg} (x^2 - 2 x+ 2 )
  +88 x^3+74 x^2+38 x+18 \biggr) \Big]
\end{aligned}
\end{align}
\begin{align}
\label{eq:PGENgg}
P_{gg}^{\text{GEN}\,(2)}(x) &= C_F n_f T_R \Big[
 \begin{aligned}[t]
& -4 (1 + x) \text{Li}_2(x) \left(a_{gq}-a_{qg}\right) +\frac{2}{3} \pi ^2 (1 + x) \left(a_{gq}-a_{qg}\right)
  -2 (1 + x) \log ^2(x) \left(c_{gq}-c_{qg}+1\right) \\
  & +\frac{2}{3} \log (x) \left(\left(4 x^2+6 x+6\right) a_{gq}-3 x a_{qg}
  - c_{gq} \left(3x + \frac{4}{x} + 6\right)
  - c_{qg} (4x^2 + 6x + 3)
  -15 x-9\right) \\
  & 
  \left(a_{gq}-a_{qg}\right)
  \frac{2 
  (1-x)(4x^2+7x+4)
  }{3 x} \log (1-x) \\
  & -\frac{2 (x-1) \left(3 x a_{gq}+\left(13 x^2+10 x+13\right) a_{qg}
  - 13 (c_{gq} + c_{qg})(x^2 + x + 1)
  -30 x^2-66 x+6\right)}{9 x} \\
  & -\delta (1-x) \Big]
\end{aligned}  \displaybreak  \\ \notag
 &\quad + C_A n_f T_R \Big[
 \begin{aligned}[t]
& \frac{4}{3} a_{gg} \left[\frac{\log (1-x)}{1-x}\right]_+
  -\frac{2}{9} \left(3 b_{gg}+10\right) \left[\frac{1}{1-x}\right]_+ \\
  & +  a_{gg} \frac{4 \left(-x^3+x^2-2 x+1\right)}{3 x} \log (1-x)
  +\frac{4 \left(\left(x^2-x+1\right)^2 c_{gg}
  + x(1-x)(1+x)
  \right)}{3 (x-1) x} \log (x) \\
  & +\frac{2 \left(23 x^3-19 x^2+29 x-23\right)}{9 x}
  -\frac{2}{3} \left(\Delta_{gg}+2\right) \delta (1-x) \Big]
\end{aligned} \\ \notag
 &\quad + C_A^2 \Big[
 \begin{aligned}[t]
& -\frac{11}{3} a_{gg} \left[\frac{\log (1-x)}{1-x}\right]_+
  +\left(\frac{11}{6}b_{gg} -\frac{\pi ^2}{3}+\frac{67}{9}\right) \left[\frac{1}{1-x}\right]_+ \\
  & -4 \text{Li}_2(-x) p_{gg}(-x) -4 \log (1 + x) \log (x) p_{gg}(-x) -4 \log (1-x) \log (x) p_{gg}(x) \\
  & -\frac{2 \left(-x^2+x+1\right)^2}{x^2-1} \log ^2(x)
  +\frac{\left(2 x^3+2 x^2+4 x+3\right)}{3 x+3} \pi ^2
  + a_{gg} \frac{11 \left(x^3-x^2+2 x-1\right)}{3 x} \log (1-x) \\
  & +\frac{\left(x 
  (1-x)(44 x^2 - 11x + 25)
  -11 \left(x^2-x+1\right)^2 c_{gg}\right)}{3 (x-1) x} \log (x) \\
  & -\frac{1}{18} (109 x + 25)
  +  \left(\frac{11}{6}\Delta_{gg} +3 \zeta_3+\frac{8}{3}\right) \delta (1-x) \Big]
\end{aligned}
\end{align}

%% file: splitfns/PDIS_formatted.tex
\begin{align}
\label{eq:PDISqiqi}
P_{q_iq_i}^{\text{DIS}\,(2)}(x) &= C_F n_f T_R \Big[
 \begin{aligned}[t]
& \frac{4}{3} \left[\frac{\log (1-x)}{1-x}\right]_+
  -\frac{29}{9} \left[\frac{1}{1-x}\right]_+
  -\frac{4}{3} \log (x) p_{qq}(x) -\frac{2}{3} (1 + x) \log (1-x) \\
  & +\frac{2}{9} (17 x+8)
  -\left(\frac{19}{6} + \frac{4 \pi ^2}{9}\right) \delta (1-x) \Big]
\end{aligned} \\ \notag
 &\quad + C_F T_R \Big[
 \begin{aligned}[t]
& \log ^2(1-x) p_{qg}(x) -2 \log (1-x) \log (x) p_{qg}(x) -3 \text{Li}_2(x)
  +\frac{1}{6} \pi ^2 \left(-4 x^2+4 x+1\right) 
  \\ &
  +\left(2 x^2-3 x-\frac{3}{2}\right) \log ^2(x) 
  +\left(-\frac{22 x^2}{3}+7 x+\frac{4}{3 x}-\frac{5}{2}\right) \log (1-x) +\left(10 x^2-4 x-\frac{1}{2}\right) \log (x) \\
  & -\frac{80 x^2}{9} +\frac{38 x}{3} +\frac{26}{9 x} -\frac{67}{6} \Big]
\end{aligned} \\ \notag
 &\quad + C_F^2 \Big[
 \begin{aligned}[t]
& -2 \log (1-x) \log (x) p_{qq}(x) - \frac{1}{2} (x + 1) \log ^2(x) \\
  & +\frac{\left(-2 x^2+2 x+3\right)}{x-1} \log (x)
  +5 (x-1)
  +\left(6 \zeta_3+\frac{3}{8}-\frac{\pi ^2}{2}\right) \delta (1-x) \Big]
\end{aligned} \\ \notag
 &\quad + C_A C_F \Big[
 \begin{aligned}[t]
& -\frac{11}{3} \left[\frac{\log (1-x)}{1-x}\right]_+
  +\left(\frac{367}{36}-\frac{\pi ^2}{3}\right) \left[\frac{1}{1-x}\right]_+
  +\frac{1}{2} \log ^2(x) p_{qq}(x) +\frac{1}{6} \pi ^2 (1 + x) \\
  & +\frac{\left(8 x^2+14\right)}{3-3 x} \log (x) +\frac{11}{6} (1 + x) \log (1-x)
  -\frac{23}{18} (11 x+2)
  +\left(-3 \zeta_3+\frac{215}{24}+\frac{11 \pi ^2}{9}\right) \delta (1-x) \Big]
\end{aligned}
\end{align}

\begin{align}
\label{eq:PDISqiqk}
P_{q_iq_k}^{\text{DIS}\,(2)}(x) = P_{q_i\overline{q}_k}^{\text{DIS}\,(2)}(x) &= C_F T_R \Big[
 \begin{aligned}[t]
& \log ^2(1-x) p_{qg}(x) -2 \log (1-x) \log (x) p_{qg}(x) -3 \text{Li}_2(x)
  +\frac{1}{6} \pi ^2 \left(-4 x^2+4 x+1\right) 
  \\ &
  +\left(2 x^2-3 x-\frac{3}{2}\right) \log ^2(x)
  +\left(-\frac{22 x^2}{3}+7 x+\frac{4}{3 x}-\frac{5}{2}\right) \log (1-x) 
  \\ &
  +\left(10 x^2-4 x-\frac{1}{2}\right) \log (x)
  -\frac{80 x^2}{9} +\frac{38 x}{3} +\frac{26}{9 x} -\frac{67}{6} \Big]
\end{aligned}
\end{align}

\begin{align}
\label{eq:PDISqiqbi}
P_{q_i\overline{q}_i}^{\text{DIS}\,(2)}(x) &= C_F T_R \Big[
 \begin{aligned}[t]
& \log ^2(1-x) p_{qg}(x) -2 \log (1-x) \log (x) p_{qg}(x) -3 \text{Li}_2(x)
  +\frac{1}{6} \pi ^2 \left(-4 x^2+4 x+1\right) 
  \\ &
  +\left(2 x^2-3 x-\frac{3}{2}\right) \log ^2(x)
  +\left(-\frac{22 x^2}{3}+7 x+\frac{4}{3 x}-\frac{5}{2}\right) \log (1-x) 
  \\ &
  +\left(10 x^2-4 x-\frac{1}{2}\right) \log (x)
  -\frac{80 x^2}{9} +\frac{38 x}{3} +\frac{26}{9 x} -\frac{67}{6} \Big]
\end{aligned} \\ \notag
 &\quad + C_F^2 \Big[
 \begin{aligned}[t]
& -4 \text{Li}_2(-x) p_{qq}(-x) -\frac{1}{3} \pi ^2 p_{qq}(-x) +\log ^2(x) p_{qq}(-x) \\
  & -4 \log (1 + x) \log (x) p_{qq}(-x)
  +2 (1 + x) \log (x)
  -4 x +4 \Big]
\end{aligned} \\ \notag
 &\quad + C_A C_F \Big[
 \begin{aligned}[t]
& 2 \text{Li}_2(-x) p_{qq}(-x) +\frac{1}{6} \pi ^2 p_{qq}(-x) -\frac{1}{2} \log ^2(x) p_{qq}(-x) \\
  & +2 \log (1 + x) \log (x) p_{qq}(-x)
  -(x + 1) \log (x)
  +2 (x-1) \Big]
\end{aligned}
\end{align}

\begin{align}
\label{eq:PDISqig}
P_{q_ig}^{\text{DIS}\,(2)}(x) &= n_f T_R^2 \Big[
 \begin{aligned}[t]
& 2 (1 + 2 x)^2 \text{Li}_2(x) -\frac{1}{3} \pi ^2 (1 + 2 x)^2 +(1 + 2 x)^2 \log ^2(x) \\
  & +\left(12 x^2-8 x-4\right) \log (1-x) +4 \left(5 x^2+8 x+1\right) \log (x)
  -59 x^2 +48 x +11 \Big]
\end{aligned} \\ \notag
 &\quad + C_F T_R \Big[
 \begin{aligned}[t]
& -2 \text{Li}_2(x) p_{qg}(x) -2 \log (1-x) \log (x) p_{qg}(x) +\left(2 x^2-x+\frac{1}{2}\right) \log ^2(x) \\
  & +\left(6 x^2-6 x-1\right) \log (1-x) +\left(-6 x^2+2 x+\frac{1}{2}\right) \log (x)
  +2 x^2 -\frac{7 x}{2} +1 \Big]
\end{aligned} \\ \notag
 &\quad + C_A T_R \Big[
 \begin{aligned}[t]
& -\text{Li}_2\left(x^2\right) p_{qg}(-x) +\log ^2(1-x) p_{qg}(x) -2 \log (x) \log (1-x) p_{qg}(x) \\
  & -2 \log (x) \log (1 + x) p_{qg}(-x) +4 (x-1) x \text{Li}_2(x) -\frac{2}{3} \pi ^2 (x-2) x -(6 x + 2) \log ^2(x) \\
  & +\left(-\frac{67 x^2}{3}+20 x+\frac{4}{3 x}-1\right) \log (1-x) +x (25 x-24) \log (x)
  +\frac{1}{18} \left(407 x^2-276 x+\frac{52}{x}-165\right) \Big]
\end{aligned}
\end{align}

\begin{align}
\label{eq:PDISgq}
P_{gq}^{\text{DIS}\,(2)}(x) &= C_F n_f T_R \Big[
 \begin{aligned}[t]
& -\frac{8}{3} \left[\frac{\log (1-x)}{1-x}\right]_+
  +2 \left[\frac{1}{1-x}\right]_+
  +4 (1 + x) \text{Li}_2(x) -\frac{2}{3} \pi ^2 (1 + x) +2 (1 + x) \log ^2(x) \\
  & +\frac{2}{3} \left(4 x^2+3 x-\frac{8}{x}+3\right) \log (1-x) +\frac{\left(8 x^3-34 x^2+24 x+10\right)}{3-3 x} \log (x) \\
  & -\frac{4 \left(24 x^3+17 x^2-31 x+13\right)}{9 x}
  +\left(6+\frac{4 \pi ^2}{9}\right) \delta (1-x) \Big]
\end{aligned} \\ \notag
 &\quad + C_F^2 \Big[
 \begin{aligned}[t]
& 3 \left[\frac{\log (1-x)}{1-x}\right]_+
  +\left(\frac{45}{4}+\frac{4 \pi ^2}{3}\right) \left[\frac{1}{1-x}\right]_+
  -6 \left[\frac{\log ^2(1-x)}{1-x}\right]{}_+ +\left(2 x+\frac{4}{x}-1\right) \text{Li}_2(x) \\
  & +\frac{\left(3 x^2+1\right)}{2 (x-1)} \log ^2(x) +\frac{\left(2 x^3+6 x^2-4 x+4\right)}{x-x^2} \log (x) \log (1-x) \\
  & -\frac{1}{6} \pi ^2 (4 x + 7) +\left(x-\frac{4}{x}+7\right) \log ^2(1-x) 
  -\frac{3}{2} (5 x+4) \log (1-x) +(6 x+4) \log (x) \\
  & -\frac{1}{2} (13 x + 23)
  +\left(\frac{27}{4}-4 \zeta_3\right) \delta (1-x) \Big]
\end{aligned} \\ \notag
 &\quad + C_A C_F \Big[
 \begin{aligned}[t]
&  6 \left[\frac{\log ^2(1-x)}{1-x}\right]_+ 
   +\frac{4}{3} \left[\frac{\log (1-x)}{1-x}\right]_+ 
   -\left(\frac{29}{2} + \frac{4 \pi ^2}{3}\right) \left[\frac{1}{1-x}\right]_+ \\
  &-\text{Li}_2\left(x^2\right) p_{gq}(-x)
  -2 \log (x) \log (1 + x) p_{gq}(-x) -\left(6 x + \frac{8}{x} + 2\right) \text{Li}_2(x) \\
  & +\frac{1}{3} \pi ^2 \left(2 x^2+x+6\right) +\left(-2 x^2+x+\frac{4}{x}-8\right) \log ^2(1-x) \\
  & +\frac{\left(2 x^3-3 x^2+2 x+1\right)}{1-x} \log ^2(x)
  +\frac{2 \left(2 x^4-4 x^3+9 x^2-7 x+4\right)}{(x-1) x} \log (1-x) \log (x) \\
  & +\left(6 x^2-2 x+\frac{4}{3 x}+6\right) \log (1-x) +\frac{\left(-26 x^3+22 x^2-30 x+47\right)}{3 (x-1)} \log (x) \\
  & +\frac{116 x^2}{9} +\frac{49 x}{9} +\frac{1}{x} +\frac{181}{9}
  +\left(4 \zeta_3-\frac{33}{2}-\frac{13 \pi ^2}{18}\right) \delta (1-x) \Big]
\end{aligned}
\end{align}

\begin{align}
\label{eq:PDISgg}
P_{gg}^{\text{DIS}\,(2)}(x) &= n_f^2 T_R^2 \Big[
 \begin{aligned}[t]
& -\frac{4}{3} \log (1-x) p_{qg}(x) +\frac{4}{3} \log (x) p_{qg}(x)
  +\frac{4}{3} \left(8 x^2-8 x+1\right) \Big]
\end{aligned} \\ \notag
 &\quad + C_F n_f T_R \Big[
 \begin{aligned}[t]
& -2 \log ^2(1-x) p_{qg}(x) +4 \log (x) \log (1-x) p_{qg}(x) +6 \text{Li}_2(x) \\
  & +\frac{1}{3} \pi ^2 \left(4 x^2-4 x-1\right) +\left(-4 x^2+2 x-1\right) \log ^2(x) \\
  & +\left(\frac{44 x^2}{3}-14 x-\frac{8}{3 x}+5\right) \log (1-x) +\left(-\frac{44 x^2}{3}+8 x-3\right) \log (x) \\
  & +\frac{1}{3} \left(36 x^2-16 x+7\right)
  -\delta (1-x) \Big]
\end{aligned} \\ \notag
 &\quad + C_A n_f T_R \Big[
 \begin{aligned}[t]
& -\frac{20}{9} \left[\frac{1}{1-x}\right]_+
  +\frac{11}{3} \log (1-x) p_{qg}(x) +\left(-\frac{22 x^2}{3}+6 x-5\right) \log (x) \\
  & +\frac{-218 x^3+226 x^2+25 x-46}{9 x}
  -\frac{4}{3} \delta (1-x) \Big]
\end{aligned} \\ \notag
 &\quad + C_A^2 \Big[
 \begin{aligned}[t]
& \left(\frac{67}{9}-\frac{\pi ^2}{3}\right) \left[\frac{1}{1-x}\right]_+
-4 \text{Li}_2(-x) p_{gg}(-x) -4 \log (1 + x) \log (x) p_{gg}(-x) 
  \\ &
  -4 \log (1-x) \log (x) p_{gg}(x) 
  -\frac{2 \left(-x^2+x+1\right)^2}{x^2-1} \log ^2(x) +\frac{\pi ^2 \left(2 x^3+2 x^2+4 x+3\right)}{3 x+3} \\
  & +\frac{1}{3} \left(-44 x^2+11 x-25\right) \log (x)
  -\frac{1}{18}(109 x + 25)
  +\left(3 \zeta_3+\frac{8}{3}\right) \delta (1-x) \Big]
\end{aligned}
\end{align}